\documentclass[aps,twocolumn,showpacs,prl,amsmath,amssymb,floatfix,superscriptaddress]{revtex4-1}

\usepackage{color}
\usepackage{bbm}
\usepackage{graphicx}
\usepackage{dcolumn}
\usepackage{bm}
\usepackage{array}
\usepackage{float}
\usepackage{supertabular}
\usepackage{longtable}
\usepackage{mathrsfs}
\usepackage{txfonts}
\usepackage{wasysym}
\usepackage[T1]{fontenc}
\usepackage[latin1]{inputenc}
\usepackage{amssymb}
\usepackage[usenames,dvipsnames]{xcolor}
\usepackage{amsmath}
\usepackage{amstext}
\usepackage{latexsym}
\usepackage[colorlinks=true,citecolor=Cerulean,linkcolor=RubineRed,urlcolor=Cerulean]{hyperref}

\graphicspath{{../images/}}

\begin{document}
\title{
 Quantum Simulation  of Generic Many-Body Open System Dynamics Using Classical Noise
 }

\author{A. Chenu}
\affiliation{Department of Chemistry, Massachusetts Institute of Technology,
77 Massachusetts Avenue, Cambridge, MA 02139, USA}
\author{M. Beau}
\affiliation{Department of Physics, University of Massachusetts, Boston, MA 02125, USA}
\author{J. Cao}
\affiliation{Department of Chemistry, Massachusetts Institute of Technology,
77 Massachusetts Avenue, Cambridge, MA 02139, USA}
\author{A. del Campo}
\affiliation{Department of Physics, University of Massachusetts, Boston, MA 02125, USA}

\newcommand{\be}{\begin{equation}}
\newcommand{\ee}{\end{equation}}
\newcommand{\bea}{\begin{eqnarray}}
\newcommand{\eea}{\end{eqnarray}}

\newcommand{\aurelia}{\color{blue}}
\newcommand{\comment}{\color{blue}}

\def\S{\mathcal{S}}
\def\E{\mathcal{E}}
\def\q{{\bf q}}

\def\G{\Gamma}
\def\L{\Lambda}
\def\la{\lambda}
\def\g{\gamma}
\def\al{\alpha}
\def\s{\sigma}
\def\e{\epsilon}
\def\k{\kappa}
\def\ve{\varepsilon}
\def\l{\left}
\def\r{\right}
\def\te{\mbox{e}}
\def\d{{\rm d}}
\def\t{{\rm t}}
\def\K{{\rm K}}
\def\N{{\rm N}}
\def\H{{\rm H}}
\def\la{\langle}
\def\ra{\rangle}
\def\om{\omega}
\def\Om{\Omega}
\def\vep{\varepsilon}
\def\wh{\widehat}
\def\tr{{\rm Tr}}
\def\da{\dagger}
\def\iz{\left}
\def\zi{\right}
\newcommand{\beq}{\begin{equation}}
\newcommand{\eeq}{\end{equation}}
\newcommand{\beqa}{\begin{eqnarray}}
\newcommand{\eeqa}{\end{eqnarray}}
\newcommand{\intf}{\int_{-\infty}^\infty}
\newcommand{\into}{\int_0^\infty}
\newcommand{\bra}[1]{\left \langle #1 \right|}
\newcommand{\ket}[1]{\left | #1 \right \rangle}
\newcommand{\der}[2] {\frac{d #1}{d #2}}
\newcommand{\parder}[2] {\frac{\delta #1}{\delta #2}}

\begin{abstract} 
We introduce a  scheme for the quantum simulation of many-body decoherence based on the unitary evolution of a stochastic Hamiltonian. Modulating the strength of the interactions with stochastic processes, we show that the noise-averaged density matrix simulates an effectively open dynamics governed by $k$-body Lindblad operators. Markovian dynamics can be accessed with white-noise fluctuations; non-Markovian dynamics requires colored noise. The time scale governing the fidelity decay under many-body decoherence is shown to scale as $N^{-2k}$ with the system size $N$. Our proposal can be readily implemented in a variety of quantum platforms including optical lattices, superconducting circuits and trapped ions.

\pacs{03.65.Yz,03.67.Bg,42.50.Dv}
\end{abstract}


\maketitle

Understanding the nonequilibrium dynamics of a quantum system embedded in an environment is a long-standing problem at the core of the foundations of physics. Environmentally induced decoherence paves the way to the emergence of classical reality from a quantum substrate. 
The decoherence program and its extensions such as quantum Darwinism are focused on it \cite{Zurek03}.
The open quantum dynamics of a system is as well of relevance to quantum technologies. 

While it is often desirable to beat decoherence and dissipation by suppressing system-environment interactions \cite{Viola99,Knill00}, new paradigms have emerged that fully embrace this coupling.  
To date, a variety of approaches have been put forward to simulate the reduced dynamics of an open quantum system \cite{Lloyd96,CZ12,Nori14}, including the engineering of quantum jump operators via digital quantum simulation \cite{Mueller11,Barreiro11}, or encoding the role of the environment in an auxiliary qubit \cite{Lloyd96,Nori11}.   
Important instances also include dissipative state preparation and quantum computation \cite{Plenio99, Plenio02, Diehl08,Kraus2008, VWC09, Zanardi2016}. 
Recent efforts focus on the possibility of engineering the environment to which the system is coupled \cite{LV01,VWC09,Boixo13}, which provides new avenues for quantum simulation of exotic phases of quantum matter \cite{Lloyd96,CZ12,Nori14}. 
Engineering of artificial baths is also motivated by the need to compute thermal averages in a variety of fields ranging from statistical mechanics \cite{Patane08,Viyuela14} to machine learning \cite{Shabani15}. Further applications include the characterization and quantification of quantum non-Markovian behavior \cite{Rivas14} and its experimental detection  \cite{NM3}.
As an alternative, one can resort to a unitary quantum circuit \cite{Kliesch11},  e.g., in combination with measurement of multi-time correlation functions  \cite{DiCandia15}, for which efficient quantum algorithms have been developed \cite{Pedernales14}.

 In this Letter, we introduce a versatile scheme for the quantum simulation of the open dynamics of a many-body system embedded in an environment to which it couples via  many-body interactions. The open-system dynamics is simulated in another, more controllable experimental platform, by adding appropriate classical noise processes.  Our scheme exploits current technologies for digital and analog quantum simulation of unitary dynamics, and can be readily implemented in various experimental platforms such as trapped ions, superconducting circuits and cold atoms.
  
\begin{figure}
\includegraphics[width= 0.8\columnwidth]{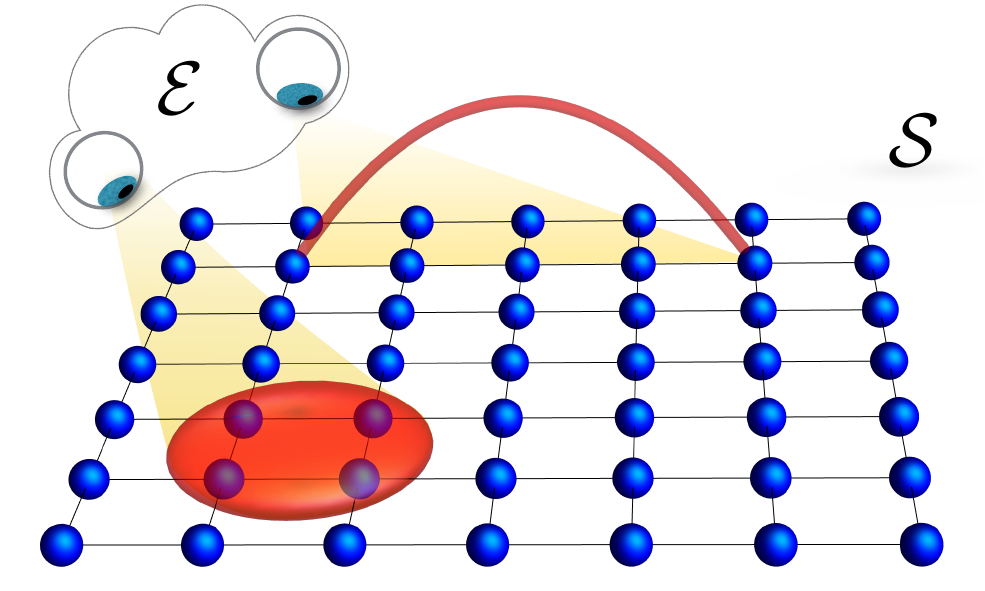}\\
\caption{{\bf Quantum simulation of many-body decoherence. } 
The implementation in a quantum simulator of the unitary dynamics generated by a Hamiltonian with  stochastic many-body terms  is used to study the open dynamics  induced by an  environment $\E$ that monitors many-body  operators of the system $\S$. The red traits illustrate the $k$-body interactions which are general in our simulation scheme -- specifically long-range and 4-body interaction in this illustration.   \label{QSMBDeco_scheme}}
\end{figure}

 Our approach is based on the quantum simulation of an isolated many-body system 
 described by a stochastic Hamiltonian, 
 where classical noise is used as a tool to simulate many-body open-system dynamics. 
 In particular, we focus on the addition of noise (understood as a stochastic modulation in time) to the coupling constants of $k$-body operators in the Hamiltonian, and show that 
 the ensemble-average over noise realizations is described by a density matrix that evolves according to 
a master equation with many-body Lindblad operators. Markovian dynamics can be accessed modulating the coupling constants with a white noise; non-Markovian dynamics requires colored noise.  The scheme is illustrated in Fig. \ref{QSMBDeco_scheme}.
We characterize the resulting many-body decoherence dynamics by identifying the time scale governing the fidelity decay.

{\it Scheme for the quantum simulation of many-body decoherence.---} The reduced dynamics of a system embedded in an environment is generally described by a master equation of the form
\beqa
\label{ME}
\frac{d}{dt}\rho(t)=-\frac{i}{\hbar} [\hat{H}_T(t),\rho(t)]+\mathcal{D}[\rho(t)],
\eeqa
where $\rho(t)$ is the reduced density matrix of a ``target'' system, with Hamiltonian $\hat{H}_T(t)$, interacting with an environment. 
The first term on the r.h.s. accounts for the unitary part of the evolution; the second term accounts for the nonunitary dynamics resulting from the interaction with the environment, which is described by the dissipator $\mathcal{D}[\cdot]$. We aim at the quantum simulation of this master equation when the $\hat{H}_T(t)$ Hamiltonian describes a many-body quantum system. We shall see that our simulation scheme, which relies on the unitary  evolution of a related stochastic simulator Hamiltonian $\hat{H}_S(t)$,
generates a family of dissipators leading to  many-body decoherence.

Specifically, our scheme utilizes the unitary dynamics of a stochastic wave function $\ket{\psi_{\rm st}(t)}$ and requires the experimental implementation of the stochastic Hamiltonian 
\beqa
\label{HSimulator}
\hat{H}_{S}(t)&=&\hat{H}_T(t)+\sum_\alpha \lambda_\alpha(t) \, \hat{L}_\alpha, 
\eeqa
in the quantum platform.  The  Hamiltonian of the quantum simulator, $\hat{H}_{S}(t)$, is composed of the target Hamiltonian, $\hat{H}_T(t)$, describing the system one aims at simulating, and a stochastic part that includes a set of operators $\hat{L}_\alpha$ with noisy coupling constants $\lambda_\alpha(t)$.   This stochastic part will be used to engineer the dissipator in (\ref{ME}) leading to many-body decoherence. 

For the sake of experimental implementation, we consider the simulator and target Hamiltonians to be Hermitian. Hermiticity carries over the stochastic term, yielding $\sum_\alpha \lambda_\alpha(t) \hat{L}_\alpha = \sum_\alpha \lambda_\alpha^*(t) \hat{L}_\alpha^\dag$. As a result, $\hat{L}_\alpha$  need not be Hermitian if the coupling constants  $\lambda_\alpha(t)$ take complex values. 
We choose the latter to be of the form $ \lambda_\alpha(t) \equiv  \hbar \sqrt{\gamma_\alpha } \, \eta_\alpha(t)$,  with  $\gamma_\alpha$ a positive real constant, and $\eta_\alpha(t)$ a complex stochastic field chosen as independent random Gaussian processes. The latter can be decomposed as $\eta_\alpha(t) = \eta'_\alpha(t) + i \eta''_\alpha(t)$, where its real $ \eta'_\alpha(t) $ and imaginary $ \eta''_\alpha(t)$ parts are two independent real Gaussian processes satisfying
\begin{equation}
\begin{split}
\la\eta'_\alpha(t)\ra = \la\eta''_\alpha(t)\ra &= \la\eta'_\alpha(t)\, \eta''_\beta(t)\ra = 0,\\
K'_{\alpha\beta}(t,t')&=\la\eta'_\alpha(t)\, \eta'_\beta(t')\ra, \\
K''_{\alpha\beta}(t,t')&=\la\eta''_\alpha(t)\, \eta''_\beta(t')\ra,
\end{split}
\end{equation}
where the bracket denotes averaging over noise realizations.
The simulator Hamiltonian (\ref{HSimulator}) can then be written in an equivalent form 
   (see \cite{SM} for details),
\beqa
\label{HS_hermitian}
\hat{H}_{S}(t)&=&\hat{H}_T(t)+ \sum_\alpha \hbar \sqrt{\gamma_\alpha} \left( \eta'_\alpha(t) \hat{A}_\alpha + \eta''_\alpha(t) \hat{B}_\alpha\right),
\eeqa
where the operators  $\hat{A}_\alpha \equiv (\hat{L}_\alpha + \hat{L}_\alpha^\dag)/2$ and $\hat{B}_\alpha \equiv i (\hat{L}_\alpha - \hat{L}_\alpha^\dag)/2$  are now Hermitian by construction, i.e. $\hat{A}_\alpha^\dag = \hat{A}_\alpha$ and $\hat{B}_\alpha ^\dag =\hat{B}_\alpha $.

The stochastic density matrix corresponding to one realization of the Gaussian processes,
$\rho_{\rm st}(t)=|\psi_{\rm st}(t)\ra\la \psi_{\rm st}(t)|$,
 is given in terms of the pure state $|\psi_{\rm st}(t)\ra$, which is obtained from the exact solution of the  Schr\"odinger equation generated by the stochastic Hamiltonian implemented in the simulator, $\hat{H}_S(t)$ in  Eq. (\ref{HS_hermitian}). 
Its time evolution is described by the stochastic quantum Liouville equation
\beqa
\begin{split}
\frac{d   \rho_{\rm st}(t)}{dt}  = &-\frac{i}{\hbar} \, [\hat{H}_T(t),\rho_{\textrm{st}}(t)]  \\
&-i \sum_\alpha \sqrt{\gamma_\alpha}\left[  \eta'_\alpha(t)  \hat{A}_\alpha + \eta''_\alpha(t) \hat{B}_\alpha, \rho_{\rm st}(t) \right] .
\end{split}
\eeqa
Averaging over different realizations of each of the stochastic processes $\{\eta_\alpha(t)\}$ leads to the noise-averaged density matrix, 
$\la \rho_{\rm st}(t) \ra =\la |\psi_\textrm{st}(t)\ra\la \psi_\textrm{st}(t)|\ra$, the dynamics of which is governed by the master equation
\begin{equation}\label{EME}
 \frac{d}{dt}\la \rho_{\rm st}(t) \ra =-\frac{i}{\hbar} [\hat{H}_T(t),\la \rho_{\rm st}(t) \ra] +\mathcal{D}[ \rho_{\rm st}(t) ],
\end{equation}
where 
\begin{equation}\label{eq:dissipator}
\mathcal{D}[ \rho_{\rm st}(t) ] = 
 - i \sum_\alpha\!\! \sqrt{\gamma_\alpha} 
\left( \left[ \hat{A}_\alpha, \la \eta'_\alpha(t) \rho_{\rm st}(t) \ra \right ] \!
+\! \left[ \hat{B}_\alpha, \la \eta''_\alpha(t) \rho_{\rm st}(t) \ra \right ] 
\right).
\end{equation}
Comparison of (\ref{EME}) with the master equation describing the reduced dynamics of open systems (\ref{ME}) enables us to identify $\mathcal{D}[\cdot]$ as a dissipator responsible for an effective nonunitary evolution of the noise-averaged density matrix. 
The explicit form of the dissipator can be evaluated using Novikov's theorem, which gives the mean value of a product of a Gaussian noise with its functional \cite{Novikov65, Budini01}. We refer the reader to \cite{SM} for the derivation that yields
\begin{eqnarray} \label{eq:D_hermitian}
\mathcal{D}[  \rho_{\rm st}(t)] &=& - \sum_{\alpha\beta} {\sqrt{\gamma_\alpha \gamma_{\beta}}}
\int_0^t dt' \\
&\times & \left( K'_{\alpha\beta}(t,t')
\left[ \hat{A}_\alpha, \la [ \hat{U}_\textrm{st}(t,t') \hat{A}_\beta \hat{U}_{\textrm{st}}^\dag(t,t'), \rho_{\textrm{st}}(t)]\ra \right] \right. \nonumber \\ 
&&\left. + K''_{\alpha\beta}(t,t')
\left[ \hat{B}_\alpha, \la [ \hat{U}_\textrm{st}(t,t') \hat{B}_\beta \hat{U}_{\textrm{st}}^\dag(t,t'), \rho_{\textrm{st}}(t)]\ra \right] \nonumber
 \right),
\end{eqnarray}
 where the time-evolution operator $\hat{U}_\textrm{st}(t,t')\equiv\mathcal{T}\exp\left[-\frac{i}{\hbar}\int_{t'}^t\hat{H}_{S}(s) ds\right]$ is defined in terms of the full stochastic Hamiltonian and $\mathcal{T}$ denotes the time-ordering operator. 

{\it Markovian limit.---}
The form of the dissipator greatly simplifies when  the  stochastic variables $\{\eta_{\alpha}(t)\}$ are described by independent white noises such that $K'_{\alpha \beta}(t,t') = K''_{\alpha \beta} (t,t') = \delta_{\alpha \beta} \delta(t-t')$.  In particular, the dissipator now only depends on the average density operator $\la \rho_{\rm st}(t) \ra$, that we hereafter denote by $\rho(t)$ to simplify the notation. Equation (\ref{eq:D_hermitian}) reduces in this case to 
\beqa \label{eq:st_WhiteNoise}
& &\mathcal{D}[  \rho(t) ]  =  -   \sum_\alpha \gamma_\alpha \left( [\hat{A}_\alpha, [\hat{A}_\alpha, \rho(t)] ] +[\hat{B}_\alpha, [\hat{B}_\alpha, \rho(t)] ]\right) \nonumber \\
& = &  \sum_{\alpha} {\gamma_\alpha} \left( \hat{L}_\alpha  \rho(t) \hat{L}_\alpha^\dag - \frac{1}{2} \{\hat{L}_\alpha^\dag \hat{L}_\alpha,  \rho(t)\} + \hat{L}_\alpha^\dag  \rho(t) \hat{L}_\alpha - \frac{1}{2} \{\hat{L}_\alpha \hat{L}_\alpha^\dag,  \rho(t) \} \right) \nonumber \\ 
& = &  \sum_{\mu} {\gamma_\mu} \left( \hat{L}_\mu \rho(t) \hat{L}_\mu^\dag - \frac{1}{2} \{\hat{L}_\mu^\dag \hat{L}_\mu,  \rho(t) \} \right),
\eeqa
where  the $\mu$ index in the last line includes the sum over the set $\{\hat{L}_\alpha\}\cup\{\hat{L}_\alpha^\dag\}$.
 This form  corresponds  to the diagonal Lindblad form \cite{Lindblad76,BP02} of a Markovian dynamics, i.e. to the form the dissipator of the reduced dynamics in (\ref{ME}) would take whenever the time scale of the system is much longer than that of the environment. 
 In this case, the equivalence between the master equations (\ref{EME}) and (\ref{ME}) and the form of the dissipator (\ref{eq:st_WhiteNoise}) shows that our scheme allows for the quantum simulation of an open system, upon identifying the noise-averaged density matrix $\la \rho_{st}(t) \ra$  with the reduced density matrix  $\rho(t)$. 
 Notice that  requiring each term in the sum to be associated with its conjugate follows from   the Hermicity of the stochastic part of the simulator Hamiltonian -- second term on the r.h.s. of Eq.~(\ref{HSimulator}). Lifting this condition would require the implementation of a non-Hermitian Hamiltonian in the simulator, which is outside the scope of our proposal since we are interested in a scheme readily implementable in current experimental platforms. 
 
Notice that, if the stochastic processes are taken to be real from the beginning ($\eta''_\alpha(t) =0$), the $\hat{L}_\alpha$ operators in (\ref{HSimulator}) then fulfill Hermiticity. The resulting dissipator
\beqa 
\label{UDiss}
\mathcal{D}[\rho(t)]=-\sum_\alpha {\gamma_\alpha} [\hat{L}_\alpha,[\hat{L}_\alpha,\rho(t)]], 
\eeqa
 becomes unital, i.e. $\mathcal{D}(\mathbb{I})=0$, where $\mathbb{I}$ is the identity operator on the Hilbert space of the target system. The noise-averaged dynamics thus leads to a monotonic decay of purity \cite{Lidar06}.

\emph{Generalization to non-Markovian dynamics.---}
While the use of white noise leads to a Lindblad dissipator simulating Markovian dynamics, many interesting processes follow a non-Markovian evolution. Such a general evolution can be obtained using colored noise. Solving the master equation (\ref{EME}) with the dissipator  (\ref{eq:D_hermitian}), although written locally in time because the dynamics generated by (\ref{HSimulator}) remains unitary,  requires the stochastic unraveling over different trajectories, or the use of perturbative schemes \cite{Budini00, Budini01}. The latter approach allows us to describe the time evolution of the density matrix  by a perturbative integro-differential equation: To second order in the strength of the noise, after approximating $\hat{U}_\textrm{st}(t,t')$ by the deterministic time-evolution operator $\hat{U}_T(t,t')\equiv \mathcal{T}\exp\left(-\frac{i}{\hbar}\int_{t'}^t\hat{H}_{T}(s) ds\right)$, Eqs. (\ref{EME})-(\ref{eq:D_hermitian}) simplify to
\beqa
&  \frac{d}{dt} \rho(t)=-\frac{i}{\hbar}[\hat{H}_T(t),& \rho(t)]\,  \\
 & -\sum_{\alpha\beta} {\sqrt{\gamma_\alpha \gamma_{\beta}}} \int_0^t dt' \Bigg( &K'_{\alpha\beta}(t,t') [\hat{A}_\alpha, [\hat{A}_{\beta}^\dag(t,t'),\rho(t)] ] \nonumber \\
\quad   &&+K''_{\alpha\beta}(t,t') [\hat{B}_\alpha, [\hat{B}_{\beta}^\dag(t,t'), \rho(t)] ] \Bigg),\nonumber
\eeqa
where $ \hat{A}_{\beta}(t,t')\equiv\hat{U}_T(t,t') \hat{A}_{\beta} \hat{U}_T^\dag(t,t')$.
A specific non-Markovian evolution can thereby be simulated from a  specific type of colored noises,  which can be designed using a filter function convoluted with a white noise signal, as in signal analysis, or via a Cholesky decomposition as described in \cite{MC13}.

{\it Many-body decoherence.---}
We next focus on a quantum simulator of $N$ particles with many-body operators $\hat{L}_\alpha$ invariant under the permutation of particles, i.e. fulfilling  
\beqa
[\hat{P},\hat{L}_\alpha]=0,
\eeqa 
where $\hat{P}$ is the permutation operator. 
Specifically, we  consider the general case of symmetric $k$-body Lindblad operators of the form 
\beqa
\label{symmL}
\hat{L}_\alpha=\sum_{i_1<\dots<i_k}{\mathbb{L}}_{i_1,\dots,i_k}^{(\alpha,k)},
\eeqa
where the sum runs over all possible tuples of $k$ particles. 
Our quantum simulation scheme then yields  a broad class of dissipators which we associate with many-body decoherence, and which  directly inherit the symmetrization over particle indices. To appreciate this, it suffices to consider the Hermitian case with a single coupling constant, taken as a real Gaussian process. Equation (\ref{UDiss}) readily gives the dissipator 
\beqa
\mathcal{D}[\rho(t)]=
-\sum_\alpha \sum_{i_1<\dots<i_k}\sum_{i'_1<\dots<i'_k}{\gamma_\alpha}[{\mathbb{L}}_{i_1,\dots,i_k}^{(\alpha,k)},[{\mathbb{L}}_{i'_1,\dots,i'_k}^{(\alpha,k)}, \rho(t)]],\:\:\: \quad 
\eeqa
The structure  of this dissipator radically differs from that customarily encountered in the study of decohering many-particle systems. Indeed,  the customary dissipators introduced in the study of decohering many-body systems result from coupling $k$ subsets of particles to independent environments, which gives rise to a single sum over the particle indices $\{i_1,\dots,i_k\}$, and is distinctly different from our result. As we shall discuss below, similar features are found in lattice systems where the symmetrization is over the lattice index. But let us first characterize the many-body dynamics. 

A natural question concerns the time scale in which many-body decoherence alters the evolution of the system. We propose the use of quantum speed limits for arbitrary physical processes \cite{Taddei13,delcampo13} to address this question. The notion of speed relies on the distance traveled during the evolution, which can be quantified by the Bures length, $\mathcal{L}[\rho(0),\rho(t)]$,  defined in terms of the fidelity between the initial and the time-evolving states. Assuming the initial state to be deterministically prepared in a pure state $|\psi(0)\ra$ at $t=0$, the fidelity simply reads  $F(t)=\la \psi(0)|\rho(t)|\psi(0)\ra=\cos^2\mathcal{L}[\rho(0),\rho(t)]$. It is well known that the short-time dynamics of the fidelity decay  follows a quadratic dependence for unitary dynamics, $F(t)= 1-|\ddot{F}(0)|t^2/2+\mathcal{O}(t^3)$,   and a linear decay for Markovian dynamics.
 Here, we recover the linear dynamics for the noise-averaged dynamics under stochastic Hamiltonians such as (\ref{HSimulator}), but with a decoherence time that now reveals a strong signature of many-body decoherence. 
For the sake of illustration, we focus on the real white-noise case, Eq.  (\ref{UDiss}). It is found that $F(t)=1 - t/\tau_D+\mathcal{O}(t^2)$, where 
\beqa \label{eq:bound}
\frac{1}{\tau_D}=\sum_\alpha\gamma_\alpha\Delta L_\alpha^2\leq \frac{1}{4}\sum_\alpha\gamma_\alpha \| \hat{L}_\alpha\|^2 ,
\eeqa
and $\Delta L_\alpha^2 =\langle \hat{L}_\alpha^2 \rangle - \langle \hat{L}_\alpha\rangle^2$.
The inequality follows from using the semi-norm of the Hermitian operator $\hat{L}_\alpha$ -- the difference between its largest and lowest eigenvalue --  as  an upper bound for the variance \cite{Boixo07}.

The seminorm of the symmetrized $k$-body Lindblad operator (\ref{symmL}) can be upper-bounded  as $
\| \hat{L}_\alpha\|\leq \sum_{i_1<\dots<i_k}\| {\mathbb{L}}_{i_1,\dots,i_k}^{(\alpha,k)}\|={N\choose k}\| {\mathbb{L}}^{(\alpha,k)}\|$, where ${N\choose k}$ is the binomial coefficient. It follows that
\beqa
\label{tauDbound}
\frac{1}{\tau_D} \leq {N\choose k}^2 \sum_\alpha\frac{\gamma_\alpha}{4} \| {\mathbb{L}}^{(\alpha,k)}\|^2 \sim  \frac{N^{2k}}{k!^2}\sum_\alpha\frac{\gamma_\alpha}{4}  \| {\mathbb{L}}^{(\alpha,k)}\|^2, 
\eeqa
i.e. the decoherence time $\tau_D$ scales as $N^{-2k}$ where $N\gg k$ is the number of particles in the quantum simulator and $k$ denotes the range of the interaction terms.  As a result, the rate of decoherence characterizing the noise-averaged dynamics generated by  $k$-body stochastic Hamiltonians with $k>1$ greatly surpasses that under local environments ($k=1$).
For the sake of illustration, we next discuss the implementation of our scheme with ultra cold atoms trapped in an optical lattice and with spin chains. 

{\it Local Lindblad operators and long-range dissipator.---}
We first consider a  Lindblad operator symmetrized over a single lattice index.
This scenario naturally arises in the quantum simulation of the Bose-Hubbard model \cite{Greiner2002}, which we use as our target Hamiltonian, taking 
\beqa \label{eq:BH}
\hat{H}_T \rightarrow \hat{H}_{\rm BH}=-J\sum_{<i,j>}\hat{b}_i^\dag\hat{b}_j+\sum_i\frac{U_i}{2}\hat{n}_i(\hat{n}_i-1),
\eeqa
where  $\hat{b}_i$ and $\hat{b}_i^\dag$  are annihilation and creation operators at site $i$,  $\hat{n}_i= \hat{b}_i^\dag\hat{b}_i$ being the site occupation number operator. The constant $J$ denotes the hopping amplitude and $U_i$ the on-site interaction. 
Such model can be implemented in an analog quantum platform formed by an optical lattice loaded with ultra cold atoms. 
In the most common setting, the interaction strength is site independent, $U_i=U$, and can be tuned via a Feshbach resonance \cite{Theis2004}. 
 It then acts as a coupling constant of an operator symmetrized over the particle index. Our scheme shows that its stochastic modulation via a single real white noise, $U \rightarrow U + 2\hbar\sqrt{\gamma}\eta(t)$,  makes the dynamics of the noise-averaged density matrix effectively open. The evolution is then dictated by the master equation (\ref{EME}) with  the dissipator
\beqa\label{eq:BH_D}
\mathcal{D}[ \rho(t) ]=-{\gamma} \sum_{i,j} [\hat{n}_i(\hat{n}_i-1),[\hat{n}_j(\hat{n}_j-1),\rho(t)].
\eeqa
While  the corresponding Lindblad operator, $\hat{L}= \sum_i\: \hat{n}_i(\hat{n}_i-1)$,  is a local one-body operator, the double sum in (\ref{eq:BH_D}) is not restricted to nearest neighbors and makes the dissipator $\mathcal{D}[\cdot]$ effectively long range. 
The obtained master equation is exact to all orders in $U$. 
Notice that such dynamics is distinctively different from a standard dissipator, that would commonly display a single sum, and could be obtained here by setting  $i=j$ in (\ref{eq:BH_D}), e.g., from the stochastic modulation of the interaction strength at each site. 
Clearly, our approach is not restricted to optical lattices and can be applied to ultracold atoms and polar molecules, including scenarios governed by three-body interactions \cite{Buchler2007a}. Nor is it restricted to local Lindblad operators, as exemplified below. 

{\it Long-range $2$-body Lindblad operators.---}
We next show how the stochastic modulation of the coupling constants in systems
with (symmetrized) two-body interactions can be used to simulate the open quantum dynamics under long-range Lindblad operators. As an example, consider the long-range Ising chain in a  transverse field $h$, 
\begin{equation}\label{LRQIM} 
\hat{H}_I = - \sum_{i<j} J_{ij} \, \sigma^z_i \sigma^z_{j} - h \sum_{i=1}^{N}\sigma^x_i.
\end{equation}
Its experimental realization has recently been reported \cite{LRIexp1,LRIexp2} with  pairwise interactions exhibiting a power-law decay $J_{ij} \propto |r_i-r_j|^{-a}$, 
as a function of the distance $r$ between two arbitrary sites $(i,j)$ of the 1D chain. 
By adding a white-noise contribution to the interactions, $J_{ij} \rightarrow J_{i j} +\hbar \sqrt{\gamma}  \eta(t)$, our results predict that the noise-averaged density matrix then obeys a master equation (\ref{EME}), where the target Hamiltonian is that of the Ising chain (\ref{LRQIM}) and the dissipator takes a many-body nonlocal  form given by 
\beqa \label{eq:ising_D}
\mathcal{D}[ \rho(t) ]=- {\gamma}\sum_{i<j}\sum_{i'<j'}\left[  \sigma^z_i \sigma^z_{j} , \left[\sigma^z_{i'} \sigma^z_{j'} ,\rho(t) \right]\right].
\eeqa
The associated dynamics is detailed in \cite{SM}.  Up to parity effects, 
the decoherence time scales quadratically with the particle number,  $\tau_D\sim 1/N^2$, for large $N$ for an initial product state. By contrast, for maximally entangled states, the bound (\ref{tauDbound}) is saturated and the enhancement scales as  $\tau_D\sim 1/N^4$, a telltale sign of many-body decoherence.
We emphasize that the 2-body long-range nature of the corresponding Lindblad operator, $\hat{L}=  \sum_{i<j}\sigma_i^z \sigma_j^z$, is directly inherited from the addition of noise to the coupling constant of the symmetrized two-body spin-spin interactions.

To summarize, we have developed a scheme for the quantum simulation of many-body decoherence,  where classical noise is a tool used to facilitate the experimental realization of such a simulation.
Our proposal relies on the unitary evolution generated by a many-body Hamiltonian that includes stochastic terms resulting from the addition of controlled noise to the interaction couplings. Averaging over the noise realizations yields an effectively open dynamics, which describes a wide variety of master equations characterized by many-body decoherence. In particular, the white-noise limit leads to Markovian dynamics, where the many-body Lindblad operators correspond to the operators introduced in the stochastic part of the simulator Hamiltonian. 
Non-Markovian effects can be accessed using colored noise. The characteristic time scale of evolution, as estimated from the fidelity decay, exhibits a strong signature of many-body decoherence as a function of the system size. 
Finally, we note that our scheme allows for the quantum simulation of a broad class of master equations that includes instances whose physical origin from first principles would be worth investigating via specific models of a system coupled to an environment.
%
Because the addition of noise in the Hamiltonian is relatively easier than engineering specific dissipations, our proposal should find broad applications in environmental engineering for quantum technologies, including dissipation-assisted  state preparation and quantum computation. Further, it  can be readily implemented in a variety of platforms, including ultracold atoms in an optical lattice, trapped ions and superconducting qubits.

\begin{acknowledgements}
{\it Acknowledgements.---}  It is a pleasure to thank W. H. Zurek for useful discussions and hospitality at Los Alamos National Laboratory during the completion of the project,  and C. Caves for useful comments on the manuscript.
We further acknowledge funding support by UMass Boston (Project No. P20150000029279) and the John Templeton Foundation,  the Swiss National Science Foundation (A.C.),  and the NSF (Grant No. CHE-1112825).
\end{acknowledgements}

\newpage

\setcounter{equation}{0}
\setcounter{figure}{0}
\setcounter{table}{0}
\makeatletter
\renewcommand{\theequation}{S\arabic{equation}}
\renewcommand{\thefigure}{S\arabic{figure}}


\begin{widetext}

\section{Supplemental material}

In this Supplemental Material, we show  how the stochastic part of the simulator Hamiltonian can be written in terms of Hermitian operators. We provide the details for the derivation of the noise-averaged density matrix master equation, which can be written in a Lindblad form. In Appendix B, we show how our scheme can as well be used to engineer $k$-body operators. Appendix C presents an application of our scheme to the long-range Ising chain. We show how the addition of classical noise to the coupling constant leads to long-time quantum revivals in the fidelity.

\section{A. Engineering master equations via Hermitian operators and real noise}

Given the general form of the  simulator Hamiltonian, $\hat{H}_S(t) = \hat{H}_T(t) + \sum_\alpha \lambda_\alpha(t) \hat{L}_\alpha$, 
Hermicity of the simulator and target Hamiltonians requires the stochastic part to satisfy $(\sum_\alpha \lambda_\alpha(t) \hat{L}_\alpha) ^\dag = \sum_\alpha \lambda_\alpha^*(t) \hat{L}_\alpha^\dag$. Using the definition of the coupling constant, $\lambda_\alpha(t) = \hbar \sqrt{\gamma_\alpha} ( \eta'_\alpha(t) + i \eta''_\alpha(t) )$, and splitting the sum into two contributions, this stochastic part can be equivalently written as  
\begin{equation*} 
\begin{split}
\sum_\alpha \lambda_\alpha(t) \hat{L}_\alpha &= \frac{1}{2} \sum_\alpha \left( \lambda_\alpha(t) \hat{L}_\alpha+ \lambda_\alpha^*(t) \hat{L}_\alpha^\dag \right) = \hbar \sum_\alpha \sqrt{\gamma_\alpha} \left( \eta'_\alpha(t) \frac{\hat{L}_\alpha + \hat{L}^\dag_\alpha}{2} + \eta''_\alpha(t) \: i \frac{\hat{L}_\alpha - \hat{L}_\alpha^\dag}{2} \right) 
=  \hbar \sum_\alpha \sqrt{\gamma_\alpha}\left( \eta'_\alpha(t) \hat{A}_\alpha + \eta''_\alpha(t) \hat{B}_\alpha \right), 
\end{split}
\end{equation*}
where $\eta'_\alpha(t)$ and $\eta''_\alpha(t)$ respectively denote the real and imaginary parts of the stochastic process $\eta_\alpha(t)$, and where we have defined $\hat{A}_\alpha \equiv (\hat{L}_\alpha + \hat{L}_\alpha^\dag)/2$ and $\hat{B}_\alpha \equiv i (\hat{L}_\alpha - \hat{L}_\alpha^\dag)/2$. We verify that, by construction, these operators are Hermitian, i.e. $\hat{A}_\alpha^\dag = \hat{A}_\alpha$ and $\hat{B}_\alpha^\dag = \hat{B}_\alpha$.
 
 \subsection{1. Derivation of the master equation for the noise-averaged density matrix }
The simulator Hamiltonian now takes the form $\hat{H}_S(t) = \hat{H}_T(t) +  \hbar \sum_\alpha \sqrt{\gamma_\alpha}\left( \eta'_\alpha(t) \hat{A}_\alpha + \eta''_\alpha(t) \hat{B}_\alpha \right)$. Considering that each operator is now Hermitian and that the stochastic processes are real,  the stochastic Liouville equation is readily given by 
\begin{equation} \label{eq:rho_stoc}
\frac{d   \rho_{\rm st}(t)}{dt}  
= -\frac{i}{\hbar} \, [\hat{H}_T(t),\rho_{\textrm{st}}(t)]  -i \sum_\alpha \sqrt{\gamma_\alpha}\left[  \eta'_\alpha(t)  \hat{A}_\alpha + \eta''_\alpha(t) \hat{B}_\alpha, \rho_{\rm st}(t) \right] .
\end{equation}
Averaging over the realizations of the noise, we obtain the dynamics for the noise-averaged density matrix, 
\beqa &\label{eq:rhost}
\frac{d \la  \rho_{\rm st}(t) \ra}{dt}
= -\frac{i}{\hbar}[\hat{H}_T(t), \la \rho_{\textrm{st}}(t) \ra] 
 - i \sum_\alpha \sqrt{\gamma_\alpha} 
\left( \left[ \hat{A}_\alpha, \la \eta'_\alpha(t) \rho_{\rm st}(t) \ra \right ] 
+ \left[ \hat{B}_\alpha, \la \eta''_\alpha(t) \rho_{\rm st}(t) \ra \right ] 
\right).
\eeqa
Since the stochastic density matrix is a functional of the stochastic fields $\eta'_\alpha(t)$ and $\eta''_\alpha(t)$, we can use Novikov's theorem to evaluate the stochastic averages in the second term on the r.h.s, which gives for these products:
\begin{equation}  \label{eq:novikov}
\begin{split}
\la \eta'_\alpha(t) \rho_{\rm st}[\eta'(t)] \ra =& \sum_\beta \int_0^t \la \eta'_\alpha(t) \eta'_\beta(t') \ra \left\la \parder{\rho_{\rm st}[\eta'(t)]}{\eta'_\beta(t')} \right \ra dt', \\
\end{split}
\end{equation}
and a similar equation for $\la \eta''_\alpha(t) \rho_{\rm st}(t) \ra$ obtained via the substitution $\eta'\rightarrow \eta''$. 

The functional derivative can be obtained solving for the stochastic density matrix from the stochastic Liouville equation (\ref{eq:rho_stoc}), which readily gives 
\begin{equation}
\rho_{\rm st}(t) = \rho_{\rm st}(t') -\frac{i}{\hbar} \int_{t'}^t ds \left[ \hat{H}_T(s) +  \sum_\alpha \hbar  \sqrt{\gamma_\alpha}\left( \eta'_\alpha(s) \hat{A}_\alpha + \eta''_\alpha(s) \hat{B}_\alpha \right), \rho_{\rm st}(s)\right].
\end{equation}
Taking the functional derivative with respect to the real part of the stochastic fields gives 
\begin{equation}  \label{eq:drho}
\begin{split}
\parder{\rho_{\rm st}[\eta'(t)]}{\eta'_\beta(t')} &= -\frac{i}{\hbar} \left[ {\hbar}  \sqrt{\gamma_\beta} \hat{A}_\beta, \rho_{\rm st}(t') \right] - \frac{i}{\hbar} \int_{t'}^t ds \left[ \hat{H}_S(s) , \parder{\rho_{\rm st}[\eta'(s)]}{\eta'_\beta(t')} \right] , 
\end{split}
\end{equation}
where we have used $\delta \eta'_\alpha(s) / \delta \eta'_\beta(t') = \delta_{\alpha \beta}\delta(s-t')$ to simplify the first term on the r.h.s. 
Taking the time derivative, we obtain a differential equation, 
\begin{equation}
\frac{d}{dt} \parder{\rho_{\rm st}[\eta'(t)]}{\eta'_\beta(t')} = -\frac{i}{\hbar} \left [\hat{H}_S(t), \parder{\rho_{\rm st}[\eta'(t)]}{\eta'_\beta(t')} \right], 
\end{equation}
similar to the stochastic Liouville equation for which the solutions are easily given using the time-evolution operator $\hat{U}_{\rm st}(t,t')$ defined in the main text and the initial condition given by the first term on the r.h.s of Eq. (\ref{eq:drho}), yielding 
\begin{equation}
\begin{split}
\parder{\rho_{\rm st}[\eta'(t)]}{\eta'_\beta(t')}  = \hat{U}_{\rm st}(t,t') \left( -i  \sqrt{\gamma_\beta} [\hat{A}_\beta, \rho_{\rm st}(t') ] \right) \hat{U}_{\rm st}^\dag(t,t'). 
\end{split}
\end{equation}
Using this expression in Eq. (\ref{eq:novikov}), we obtain the first noise-averaged product as 
\begin{equation}
\la \eta'_\alpha(t) \rho_{\rm st}[\eta'(t)] \ra = -i \sum_\beta  \sqrt{\gamma_\beta} \int_0^t dt' K'_{\alpha \beta}(t,t') \left \la [\hat{U}_{\rm st}(t,t') \hat{A}_\beta \hat{U}^\dag_{\rm st}(t,t'), \rho_{\rm st}(t)]\right \ra.
\end{equation}
Following a similar procedure for the noises forming the imaginary part of the stochastic field, we find
\begin{equation}
\la \eta''_\alpha(t) \rho_{\rm st}[\eta''(t)] \ra = -i \sum_\beta \sqrt{\gamma_\beta} \int_0^t dt' K''_{\alpha \beta}(t,t') \left \la [\hat{U}_{\rm st}(t,t') \hat{B}_\beta \hat{U}^\dag_{\rm st}(t,t'), \rho_{\rm st}(t)]\right \ra.
\end{equation}
Using these results in Eq. (\ref{eq:rhost}), we recover the master equation for the noise-averaged density matrix given in the main text [Eqs. (\ref{EME})-(\ref{eq:D_hermitian})]. 

\subsection{2. Engineering Lindblad master equations via Hermitian operators and real noise}

For the sake of illustration, we elaborate  on the Markovian case arising when the real Gaussian processes are white noises.  We allow the amplitudes of each stochastic processes to be different and define the real positive constants $\gamma'_\alpha $ and $\gamma''_\alpha$ according to 
\begin{equation}
\lambda_\alpha(t) = \hbar \sqrt{\gamma'_\alpha} \eta'_\alpha(t) + i  \hbar \sqrt{\gamma''_\alpha} \eta''_\alpha(t).
\end{equation}
Writing $\lambda_\alpha(t) $ from its real  and imaginary parts, $\lambda'_\alpha(t) $ and $\lambda''_\alpha(t)$ respectively, the stochastic processes fulfill  
\beqa
& \la \lambda'_\alpha(t)\ra=\la \lambda''_\alpha(t)\ra=0, \\
&\la \lambda'_\alpha(t)\lambda'_\beta(t')\ra=\hbar\gamma'_\alpha \delta_{\alpha\beta}\delta(t-t'),\quad \la \lambda''_\beta(t)\lambda''_\alpha(t')\ra=\hbar\gamma''_\alpha \delta_{\alpha\beta} \delta(t-t'). \nonumber
\eeqa
An explicit calculation of the dissipator describing the noise-averaged dynamics yields
\beqa
\mathcal{D}[ \rho_{\rm st}(t) ]&=&\sum_{\alpha}(\gamma_{\alpha}'-\gamma_{\alpha}'')
\left[\hat{L}_{\alpha}\la \rho_{\rm st}(t) \ra \hat{L}_{\alpha}+\hat{L}_{\alpha}^\dag \la \rho_{\rm st}(t) \ra \hat{L}_{\alpha}^\dag-\frac{1}{2}\{\hat{L}_{\alpha}\hat{L}_{\alpha}+\hat{L}_{\alpha}^\dag \hat{L}_{\alpha}^\dag, \la \rho_{\rm st}(t) \ra \}\right] \\
& +& \sum_{\alpha}(\gamma_{\alpha}'+\gamma_{\alpha}'')
\left[\hat{L}_{\alpha}^\dag \la \rho_{\rm st}(t) \ra \hat{L}_{\alpha}+\hat{L}_{\alpha} \la \rho_{\rm st}(t) \ra \hat{L}_{\alpha}^\dag-\frac{1}{2}\{\hat{L}_{\alpha}^\dag \hat{L}_{\alpha}+\hat{L}_{\alpha} \hat{L}_{\alpha}^\dag, \la \rho_{\rm st}(t) \ra \}\right].\nonumber
\eeqa
By setting egual amplitudes, i.e.  $\gamma_{\alpha}'=\gamma_{\alpha}''= \gamma_\alpha$, we obtain a dissipator in the Lindblad form
\beqa 
\mathcal{D}[ \rho_{\rm st}(t) ]=\sum_{\mu} {\gamma_\mu} \left( \hat{L}_\mu \la \rho_{\rm st}(t) \ra \hat{L}_\mu^\dag - \frac{1}{2} \left\{\hat{L}_\mu^\dag \hat{L}_\mu, \la \rho_{\rm st}(t) \ra\right\} \right),
\eeqa
where  the $\mu$ index in the last equation includes the sum over the set $\{\hat{L}_\alpha\}\cup\{\hat{L}_\alpha^\dag\}$, as given in the main text [Eq. (\ref{eq:st_WhiteNoise})].

\section{B. Quantum simulation of master equations with  $k$-body Lindblad operators.}

Our simulation scheme is generally free from errors associated with time-discretization as it is  applicable to analog, digital and hybrid simulation approaches. Nonetheless, we next illustrate how our scheme would perform in combination with the digital quantum simulation of $k$-body Hamiltonians that is amenable to  various quantum platforms, including trapped ions \cite{Mueller11,Casanova12} and superconducting circuits \cite{Mezzacapo14}. 
 In the digital approach, the evolution operator generated by the stochastic many-body simulator Hamiltonian $\hat{H}_S$ can be decomposed as the product of nonlocal spin operators via the Trotter-Suzuki decomposition \cite{Lloyd96}.  Let us consider the case in which the particles available in the simulator are qubits.
 Combining the single-site addressability of each constituent qubit with the use of entangling gates, non-local $k$-body Hamiltonian can be implemented via the sequential operation of an entangling  M\o lmer-S\o rensen, a local gate acting on one of the qubits, and an inverse $k$-body M\o lmer-S\o rensen gate \cite{Mueller11,Casanova12,Laura16}. 
As an example, given a Hamiltonian $\hat{H}_S=\sum_\ell\hat{h}_\ell$, the time evolution operator can be approximated as $\exp(-i\hat{H}_St/\hbar)\approx[\prod_\ell\exp(-i\hat{h}_\ell t/\hbar M)]^M$ for large $M$, with a discretization error that scales as $(t^2/\hbar^2M)\sum_{\ell,m}[\hat{h}_\ell,\hat{h}_m]$. 
The following exponential then takes the form 
\beqa
\label{MSU}
e^{-i\frac{gt}{\hbar}\sigma_1^z\otimes\sigma_2^x\otimes\cdots\otimes\sigma_k^x}=\hat{U}_{\rm MS}(-\pi/2,0)e^{-i\frac{gt}{\hbar}\sigma_1^z} \hat{U}_{\rm MS}(\pi/2,0),
\eeqa
where  $g$ is a generic coupling constant, $\hat{U}_{\rm MS}(\theta,\phi)=\exp[-i\theta(\hat{S}_x\cos \phi+\hat{S}_y\sin \phi)^2/4]$, the global spin operators read 
$\hat{S}_\chi=\sum_{i=1}^k\sigma_i^\chi$ with $\chi=x,y$, and $k$ is taken odd for simplicity.
Adding a stochastic fluctuation to the local single-qubit rotation so that $g\rightarrow g+ \hbar\sqrt{\gamma}\eta(t)$, our scheme predicts 
the noise-averaged dynamics to be described by a dissipator of the form
\beqa
\mathcal{D}[ \rho_{\rm st}(t) ]=-{\gamma}  [\sigma_1^z\otimes\sigma_2^x\otimes\cdots\otimes\sigma_k^x,[\sigma_1^z\otimes\sigma_2^x\otimes\cdots\otimes\sigma_k^x,\la\rho_{\rm st}(t)\ra].
\eeqa
Not surprisingly, this illustrates that master equations including  $k$-body Lindblad operators can be engineered using stochastic $k$-body Hamiltonians, which can be experimentally implemented in a digital quantum simulator. In this sense, our scheme is complementary to that in Ref \cite{Mueller11}, which allows simulation of general Lindblad dynamics with $k$-body operators using an additional qubit as an ancilla. Notice that, as illustrated with the Bose-Hubbard model and long-range Ising chain described in the main text, our scheme does not necessarily require the implementation of $k$-body operators.

\section{C. Long-range Ising chain with zero-magnetic field}\label{SM:Long-range Ising}

In this section, we study the dynamics of the long-range Ising model with zero-magnetic field, i.e. $h=0$, in the presence of a stochastic real white-noise for the two-body interaction coupling $J_{ij}\rightarrow J_{ij}+\hbar\sqrt{\gamma}\eta(t)$. To simplify reading, we write the noise-averaged density matrix as $\la \rho_{\rm st}(t)\ra \rightarrow \rho(t)$, and omit the hats on  operators. We have seen in the main body of the paper that the noise-averaged density matrix satisfies the master equation ({\it cf} Eqs. (\ref{EME}) and (\ref{eq:ising_D}) in the main text)
\begin{equation}\label{SM:MasterEqIsingCompact}
\dot{\rho}(t)=-\frac{i}{\hbar}\left[H_I,\rho(t)\right]-\frac{\gamma}{2}\left[ L, [L, \rho(t)] \right]\ ,
\end{equation}
where the long-range Ising Hamiltonian is $H_I=-\sum_{i<j}J_{ij}\sigma_i^z\sigma_j^z$ and we have defined the symmetrized Lindblad operator $L=\sum_{i<j}\sigma_i^z\sigma_j^z$. 
We first solve the dynamics, and thus discuss the effect of the range of the coupling $J_{ij}$ and of the number of spins.

We consider a chain of $N$ spins, and construct a basis $\{\ket{e_p^{(m)}}\}$ from the $2^N$ combinations of spin up and down $|a_1 a_2\cdots a_N \ra = \bigotimes_{j=1}^{N}|a_j\ra $, where the $|a_j\ra$ are eigenstates of the single Pauli matrices $\sigma_j^z$, i.e. $\sigma_j^z|a_j\ra = a_j|a_j\ra$ with $a_j=\pm 1$.
 The basis vectors are constructed from  
 $$|e_p^{(m)}\ra=\pi^{(m)} |v_p\ra \ ,$$
i.e. from the $\pi^{(m)}$ permutations in the $S_N$ symmetric group of the vector
$$
|v_p\ra=|\underbrace{-1,\dots, -1,}_{N-p\ \text{times}}\underbrace{+1,\dots, +1}_{p\ \text{times}} \ra\ ,\ p\in\{0,\dots,N\}.
$$
The label $m$ goes from $1$ to ${N \choose p}$. By convention $\pi^{(1)}$ denotes the identity so that $|e_p^{(1)}\ra = |v_p\ra$. After summing over all possible $p$ and $m$, we obtain the correct number of $2^N=\sum_{p=0}^{N}{N \choose p}$ orthonormal vectors in the basis.\\

\fbox{\begin{minipage}{50em}
\textit{Example 1: orthonormal basis for $N=4$} \\
We first define the $4+1=5$ vectors
$$|v_0\ra =| - - - - \ra\ ;\ |v_1\ra =| - - - + \ra\ ;\ |v_2\ra =| - - + + \ra;\ |v_3\ra =| - + + + \ra;\ |v_4\ra =| + + + + \ra\ ,$$
that describe the first $5$ vectors of the basis $|e_p^{(1)}\ra = |v_p\ra,\ p=\{0,1,2,3,4\}$.
Then, we construct the other $2^4-5=11$ vectors by permutations
$$|e_1^{(2)}\ra =| - - + - \ra\ ;\ |e_1^{(3)}\ra=| - + - - \ra\ ;\ |e_1^{(4)}\ra=| + - - - \ra\ ,$$
$$|e_2^{(2)}\ra =| - + - + \ra\ ;\ |e_2^{(3)}\ra=| + - - + \ra\ ;\ |e_2^{(4)}\ra=| - + + - \ra\ ;\ |e_2^{(5)}\ra=| + - + - \ra\ ;\ |e_2^{(6)}\ra=| + + - - \ra\ ,$$
$$|e_3^{(2)}\ra =| + + - + \ra\ ;\ |e_3^{(3)}\ra=| + - + + \ra\ ;\ |e_3^{(4)}\ra=| - + + + \ra\ .$$
From above, we find $\{1$, $4$, $6$, $4$, $1\}$ states for $p=\{0,\ 1,\ 2,\ 3,\ 4\}$ respectively, which in total gives $1+4+6+4+1=2^4$ states.
\end{minipage}}
\linebreak\\ 

In this representation, the density matrix $\rho(t)$ is a $2^N\times 2^N$ square matrix. As the Pauli matrices are diagonal in the basis $|e_{p}^{(m)}\ra$, we find that the Hamiltonian $H_I$ and the Lindblad operator $L$, have a diagonal form  
\begin{equation}
H_I=\text{Diag}\left(\epsilon_1,\cdots,\epsilon_{2^N}\right)\ ,\ L=\text{Diag}\left(l_1,\cdots,l_{2^N}\right)
\end{equation}  
where $\epsilon_I,\ I=1,\dots, 2^N$, and $l_I,\ I=1,\dots, 2^N$ are the eigenvalues of $H_I$ and $L$, respectively. Some of these eigenvalues can be degenerated or equal to zero. We will give more details in what follows. In general, the $\epsilon_I$ depend on the coupling $J_{ij}$ and the $l_I$'s are constant. 
Now, writing Eq. \eqref{SM:MasterEqIsingCompact} for the matrix elements $\rho_{IJ}(t)=\la e_I|\rho(t)|e_J \ra=\la e_i^{(m)}|\rho(t)|e_j^{(m)} \ra$ of the density matrix, where the $|e_I \ra= |e_{i}^{(m)}\ra$'s are vectors in the basis described above with the super-indice $I = \{i,m\}$, we find 
\begin{equation}
\dot{\rho}_{IJ}(t)=\left(-\frac{i}{\hbar}(\epsilon_I-\epsilon_J)-\frac{\gamma}{2}(l_I-l_J)^2\right)\rho_{IJ}(t)\ ,
\end{equation}   
leading to
\begin{equation}\label{SM:rhot}
\rho_{IJ}(t) = \rho_{IJ}(0)e^{-\frac{i}{\hbar}(\epsilon_I-\epsilon_J)t-\frac{\gamma}{2}(l_I-l_J)^2t}\ .
\end{equation}

Assuming the initial state is pure, $\rho(0)^2=\rho(0)$,  direct computation of the purity gives 
\begin{equation}\label{SM:purity}
p(t)\equiv \text{Tr}\left(\rho(t)^2\right)=\sum_{I,J}|\rho_{IJ}(t)|^2=\sum_{I,J}|\rho_{IJ}(0)|^2 e^{-\gamma(l_I-l_J)^2t}=1-2\sum_{I<J}|\rho_{IJ}(0)|^2\left(1-e^{-\gamma(l_I-l_J)^2t}\right)\ ,
\end{equation}
which is a decreasing function consistently with the fact that the dissipator is unital \cite{Lidar06}.   
Similarly, we can give an explicit expression of the fidelity
\begin{equation}\label{SM:fidelity}
F(t)\equiv \text{Tr}\left(\rho(t)\rho(0)\right) = \sum_{I,J}\rho_{IJ}(t)\rho_{JI}(0) = \sum_{I,J}|\rho_{IJ}(0)|^2 e^{-\frac{i}{\hbar}(\epsilon_I-\epsilon_J)t-\frac{\gamma}{2}(l_I-l_J)^2t} = 1-2\sum_{I<J}|\rho_{IJ}(0)|^2\left(1-e^{-\frac{\gamma}{2}(l_I-l_J)^2t}\cos{\left(\frac{\epsilon_I-\epsilon_J}{\hbar}t\right)}\right)\ .
\end{equation}
  
We next discuss the degeneracy of the spectrum of the symmetrized Lindblad operator $L$ and of the Hamiltonian $H_I$, starting with $L$. Because of the symmetry of the Lindblad operator $L=\sum_{i<j}\sigma_i^z\sigma_j^z$, we clearly have, for  $p\geq 0$, 
$$L| e_p^{(m)}\ra =L| e_p^{(1)}\ra=l_p| e_p^{(1)}\ra\ , \forall m\in \{1,\cdots,{N \choose p}\} .$$    
Hence the eigenvalue $l_p$ has a degeneracy equal to ${N \choose p}$ and $L$ has the representation
$$L=\text{Diag}\left(\underbrace{l_0}_{1\ \text{times}},\dots,\underbrace{l_p\dots l_p}_{{N \choose p}\ \text{times}},\dots,\underbrace{l_N}_{1\ \text{times}}\right).$$
To compute the eigenvalues $l_p$ explicitly, it suffices first, to notice that 
\begin{equation} \label{SM:Spectsigmaij}
\sigma_i^z\sigma_j^z | e_p^{(1)}\ra = \left\{ \begin{array}{ll} +1,\ \text{if}\ \{i,j\} \in \{1,N-p\} \, {\rm   or   } \,  \{N-p+1,N\} \ , \\ -1,\ \text{otherwise}\ .\end{array} \right.
\end{equation}
and second, to compute the number of terms in the expression of $L$ in each case 
\begin{gather}
\text{Number}\left(\sigma_i^z\sigma_j^z\ \text{with}\ i \ \text{and}\ j\leq\ \text{or}\ \geq N-p\right)={N-p \choose 2}+{p \choose 2}\ ,\\
\text{Number}\left(\sigma_i^z\sigma_j^z\ \text{with}\ i \ \text{or}\ j\geq N-p\right)={N \choose 2}-{N-p \choose 2}-{p \choose 2}\ .
\end{gather}
This yields the expression of $l_p$
$$l_p = 2{N-p \choose 2} + 2{p \choose 2} - {N \choose 2} = \frac{(N-2p)^2-N}{2}\ ,$$
where the maximum $l_M$ and minimum $l_m$ eigenvalues are
$$ l_M = {N \choose 2}=\frac{N(N-1)}{2}\ ,\ l_m = 4 {\left[ \frac{N}{2}\right] \choose 2}-{N \choose 2}=\left\{ \begin{array}{ll}-\frac{N}{2} \ \ \ , \ N\ \text{even} \\ \\ -\frac{N-1}{2} ,\ N\ \text{odd} \end{array} \right.\ , $$
where $ \left[ \cdot\right] $ is either the floor function $\lfloor \cdot\rfloor$  or the ceiling function $\lceil \cdot\rceil$. Therefore the degeneracy of the minimum eigenvalue depends on the parity of $N$. In turn, the degeneracy of the maximum eigenvalue is always 2, since $l_M$ is associated with the two states $| e_0 \ra = |-1, \dots,  -1\ra$ and  $| e_N \ra = |+1,\dots, +1\ra$. 

Notice that the state 
$$
|v_{\lfloor\frac{N}{2}\rfloor}\ra=|\underbrace{-1\cdots -1}_{N-\lfloor\frac{N}{2}\rfloor\ \text{times}}\underbrace{+1\cdots +1}_{\lfloor\frac{N}{2}\rfloor\ \text{times}} \ra\ ,
$$
is an eigenstate of $L$ with  minimum eigenvalue. As we discussed above, one can construct ${N\choose \lfloor\frac{N}{2}\rfloor}$ states $| e_{\lfloor\frac{N}{2}\rfloor}^{(m)}\ra $ with the same eigenvalue, and ${N\choose \lceil\frac{N}{2}\rceil}$ additional states $| e_{\lceil\frac{N}{2}\rceil}^{(m)}\ra $  if $N$ is odd. 

For example, this shows that the initial pure state $\rho_\psi=\ket{\psi}\bra{\psi}$, composed by the cat state $\ket{\psi} = (| e_0 \ra + | e_N \ra )/\sqrt{2}$, is an eigenstate state of the Lindbladian operator as $L\ket{\psi} = l_M\ket{\psi}$ and so $L\rho_\psi L-\{L^2,\rho_\psi\}/2 = 0$, yielding $[H,\rho_\psi]$=0.  
Therefore, such a state is not suitable for maximizing the decoherence. However, one can construct the  state 
\begin{equation}\label{SM:rhoM} 
 \rho_{M} = | \varphi\ra \la \varphi|,\ \text{with}\ | \varphi\ra
 = | e_0 \ra + | v_{\lfloor\frac{N}{2}\rfloor}\ra\ , 
\end{equation} 
that maximizes the value of $\Delta_\rho L^2=\la L^2\ra_\rho - \la L\ra_\rho^2$ with respect to the state $\rho$ (where $\la A\ra_\rho\equiv \text{Tr}(A\rho)$)
$$ \Delta_{\rho_M} L^2 = \frac{1}{4}\| L\|^2 = \frac{1}{4}\left(l_M - l_m\right)^2=\left\{ \begin{array}{ll}\frac{N^4}{16} \ \ \ , \ N\ \text{even}\  \\ \\ \frac{(N+1)^4}{16} ,\ N\ \text{odd}\  \end{array} \right.\ , $$ 
where $\| A\|$ denotes the seminorm of an operator $A$ defined as the difference between its largest and smallest eigenvalues. Notice that any other eigenstate $| e_{a} \ra + | e_{\lfloor\frac{N}{2}\rfloor}^{(m)}\ra$, where $a=0$ or $1$ and any $m$ label, is also associated with the smallest eigenvalue of $L$. For large $N$ we find that $\Delta_{\rho_M} L^2 \sim N^4/16 $
which agrees with the bound given in Eq. (\ref{tauDbound}) in the main text, where we find that the decoherence time for this maximum eigenstate scales as $\tau_D \sim N^{-4}$.  
The purity and fidelity of such state is simply given by
\begin{subequations}\label{SM:MaxStates}
\begin{equation}\label{SM:MaxStatesPurity}
p(t)=\frac{1}{2}+\frac{1}{2}e^{-\gamma  \|L\|^2 t}\ ,
\end{equation}
\begin{equation}\label{SM:MaxStatesFidelity}
F(t)=\frac{1}{2}+\frac{1}{2}e^{-\frac{\gamma}{2} \|L\|^2 t}\cos{\left(\frac{E}{\hbar}t\right)}\ ,
\end{equation}
\end{subequations}
where the seminorm $\|L\|=l_M-l_m$ and $E=\epsilon_{0}-\epsilon_{\lfloor \frac{N}{2} \rfloor}$ is the difference between the energy (i.e., eigenvalue of $\hat{H}_I$) of the two states $|e_{0}\ra$ and $|v_{\lfloor \frac{N}{2} \rfloor}\ra$. 
Both functions tend to one half in the limit $t\rightarrow +\infty$ which means that the coherence is suppressed at long time and the fixed point satisfying $\mathcal{L}(\rho_\infty)=0$ is diagonal in the basis $|e_{0}\ra$ and $|v_{\lfloor \frac{N}{2} \rfloor}\ra$ with equal weights, $\rho_{\infty}=\frac{1}{2}\left(|e_{0}\ra\la e_{0}|+|v_{\lfloor \frac{N}{2} \rfloor}\ra \la v_{\lfloor \frac{N}{2} \rfloor}|\right)$. At short time we find the fidelity reads $F(t)\approx 1-\gamma\Delta_{\rho_M} L^2 t$, consistently with Eq.  (\ref{eq:bound}) in the main text. \\    

To compute the spectrum of the Hamiltonian $H_I$, we use Eq.  \eqref{SM:Spectsigmaij} to find
$$ H_I| e_{p}^{(1)}\ra = \sum_{i<j} J_{ij}\sigma_i^z\sigma_j^z| e_{p}^{(1)}\ra 
=\sum_{i<j} \xi_{ij} J_{ij}| e_{p}^{(1)}\ra \ , $$
where $\xi_{ij}=\text{sgn}\left[i-(N-p)\right]\text{sgn}\left[j-(N-p)\right]$ can be equal to $\pm 1$. 
This yields to 
$$ H_I| e_{p}^{(m)}\ra = \sum_{i<j} \xi_{\pi^{(m)}(ij)} J_{ij}| e_{p}^{(m)}\ra \ , $$
where $\xi_{\pi^{(m)}(ij)}$ denotes the permutation of the signs $\xi_{ij}$. 
Thus the energy eigenvalues are $ \epsilon_p^{(m)}= \sum_{i<j} \xi_{\pi^{(m)}(ij)} J_{ij}$. 
\\

\fbox{\begin{minipage}{50em}
\textit{Example 2: spectrum of $H_I$ for $N=4$ .} \\
To illustrate the previous result, let us compute $H_I|e_2^{(1)}\ra$ with $|e_2^{(1)}\ra=|0011\ra$
$$ H_I|e_2^{(1)}\ra = +J_{12} - J_{13} - J_{14} - J_{23} - J_{24} + J_{34}\ , $$
which gives $\xi_{12}=+1,\ \xi_{13}=-1,\ \xi_{14}=-1,\ \xi_{23}=-1,\ \xi_{24}=-1,\ \xi_{34}=+1$.\\
Now we construct the state $|e_2^{(2)}\ra = \pi(|e_2^{(1)}\ra) = |0101\ra$, where $\pi$ permutes the particle $2$ and $3$. Then we have
$$ H_I|e_2^{(2)}\ra = -J_{12} + J_{13} - J_{14} - J_{23} + J_{24} - J_{34}\ , $$
where $\xi_{12}'=\xi_{13}=-1,\ \xi_{13}'=\xi_{12}=+1,\ \xi_{14}'=\xi_{14}=-1,\ \xi_{23}'=\xi_{23}=-1,\ \xi_{24}'=\xi_{34}=+1,\ \xi_{34}'=\xi_{24}=-1$, with $\xi_{ij}'=\xi_{\pi(ij)}$. 
 
\end{minipage}}
\linebreak\\ \linebreak\\ 

Interestingly, the degeneracy of the eigenvalues of the symmetrized Lindblad operator $L$ implies that the master equation \eqref{SM:MasterEqIsingCompact} does not suppress all the coherence of the initial state. Indeed, from equation \eqref{SM:rhot}, one finds that for all the $(i,j)$ for which $l_i=l_p$ exponential decay cancels. Hence the asymptotic of the density matrix for $t\rightarrow +\infty$ is
\begin{equation}\label{SM:rhotInfty}
\rho_{IJ}(t) \approx  \left\{ \begin{array}{ll} \rho_{IJ}(0)e^{-\frac{i}{\hbar}(\epsilon_I-\epsilon_J)t},\ \text{if}\ l_I=l_J, i.e., { } l_i=l_j\ , \\ 0,\ \text{otherwise}\ ,\end{array} \right.
\end{equation}
and of the fidelity is
\begin{equation}\label{SM:fidelityInfty}
F(t)\approx 1-2\sum_{I<J:l_I=l_J}|\rho_{IJ}(0)|^2\left(1-\cos{\left(\frac{\epsilon_I-\epsilon_J}{\hbar}t\right)}\right)\ ,
\end{equation}
which oscillates between two values. Therefore, in general at large time the system has some coherence with multi-frequency oscillation. 

\begin{figure}[t]
\begin{center}
\includegraphics[width=0.3\linewidth]{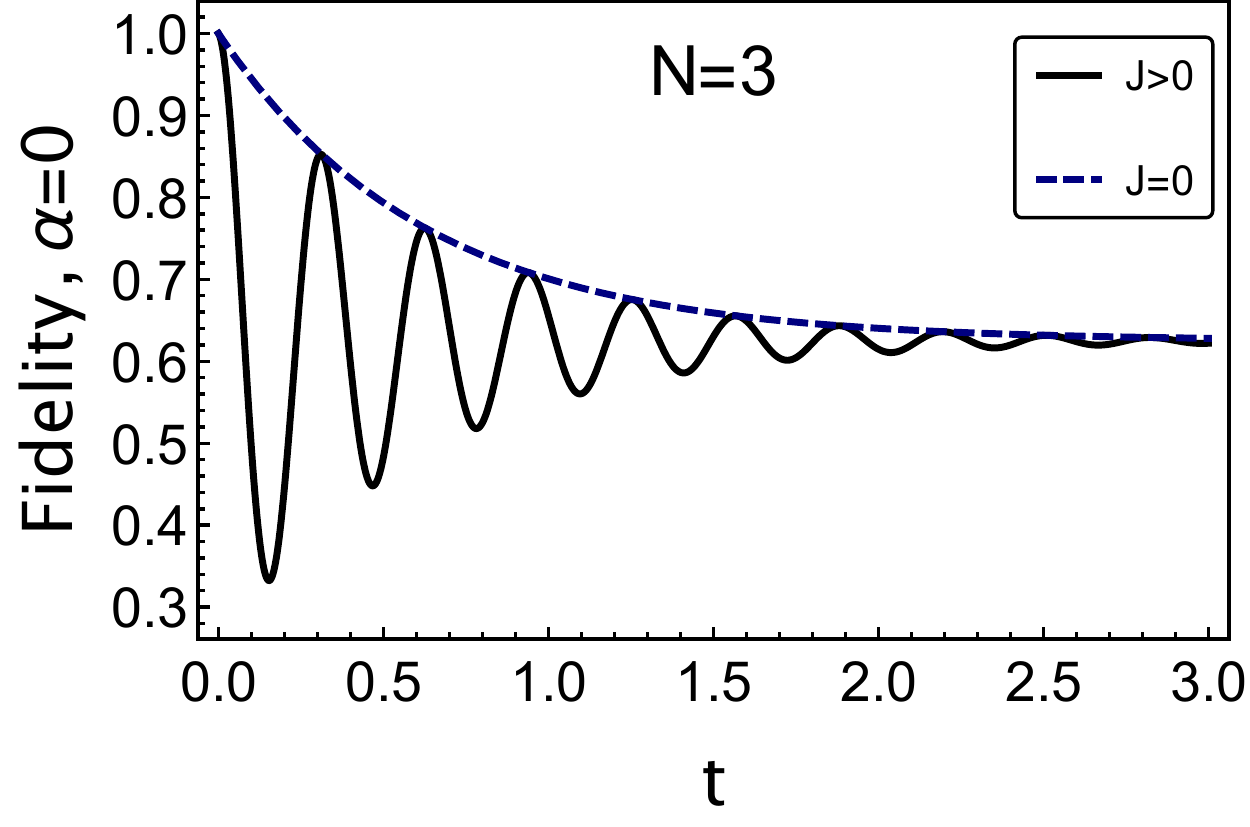}\ \ \ \includegraphics[width=0.3\linewidth]{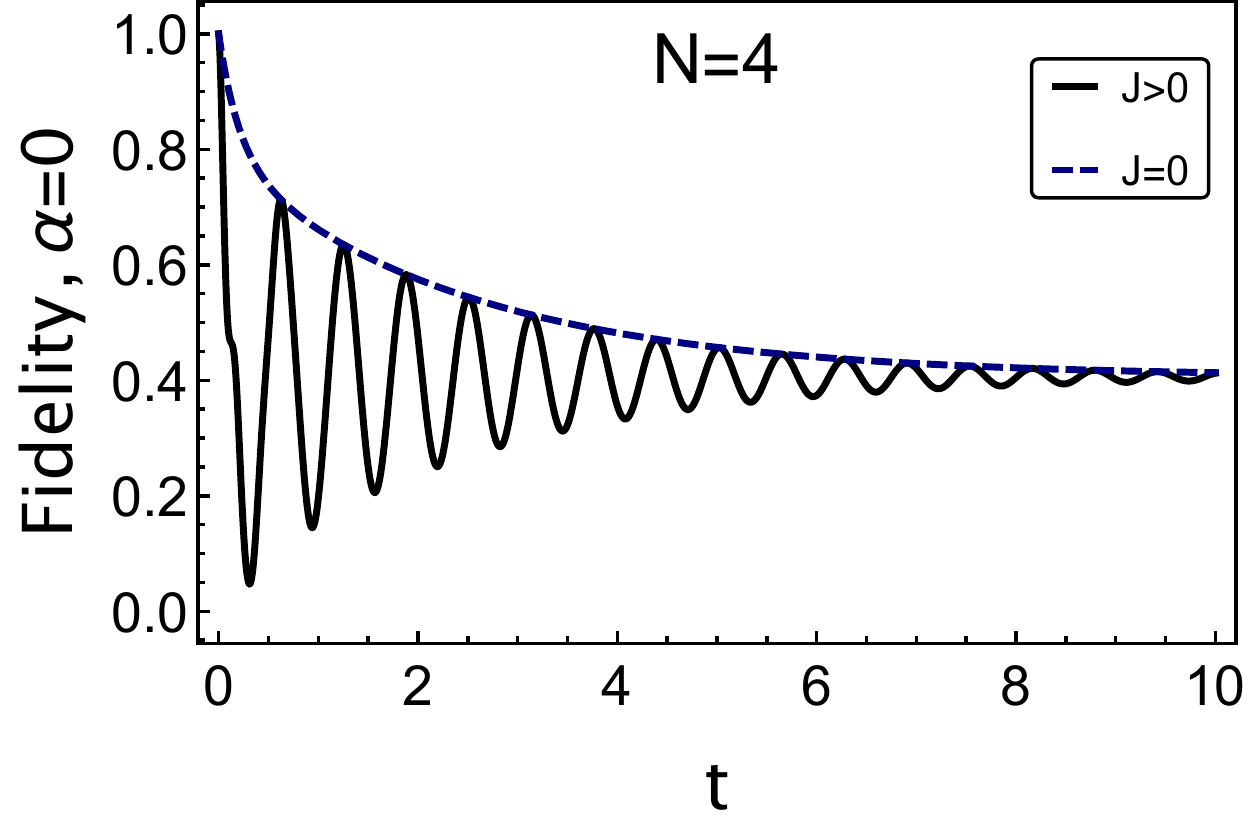}\ \ \ \includegraphics[width=0.3\linewidth]{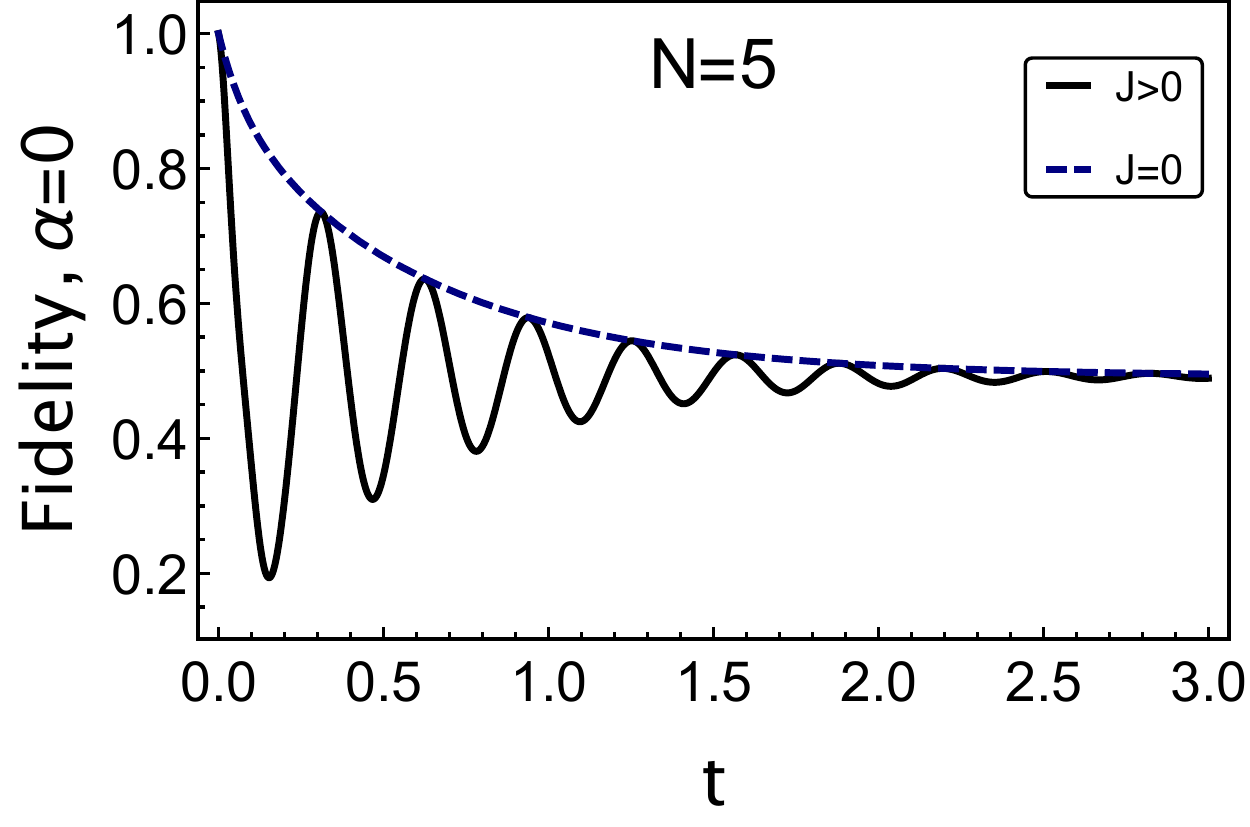}
\includegraphics[width=0.3\linewidth]{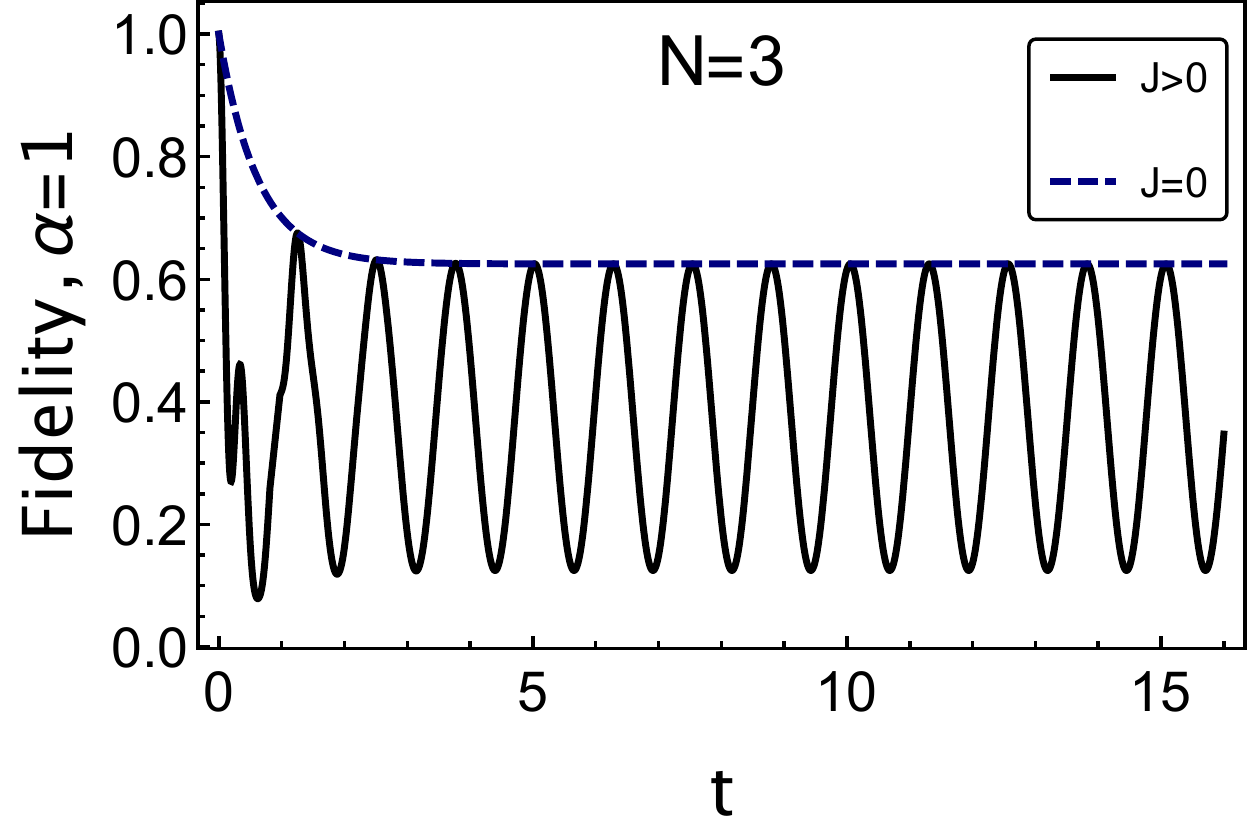}\ \ \ \includegraphics[width=0.3\linewidth]{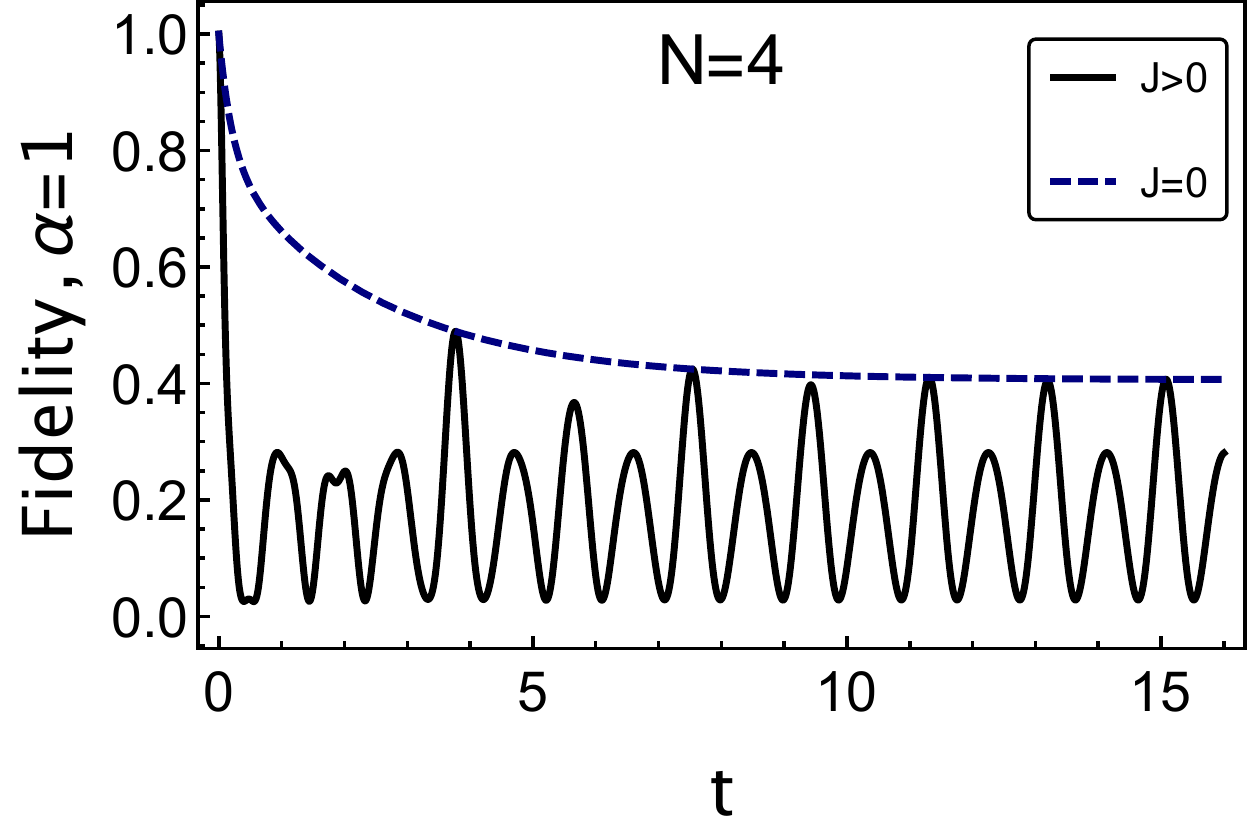}\ \ \ \includegraphics[width=0.3\linewidth]{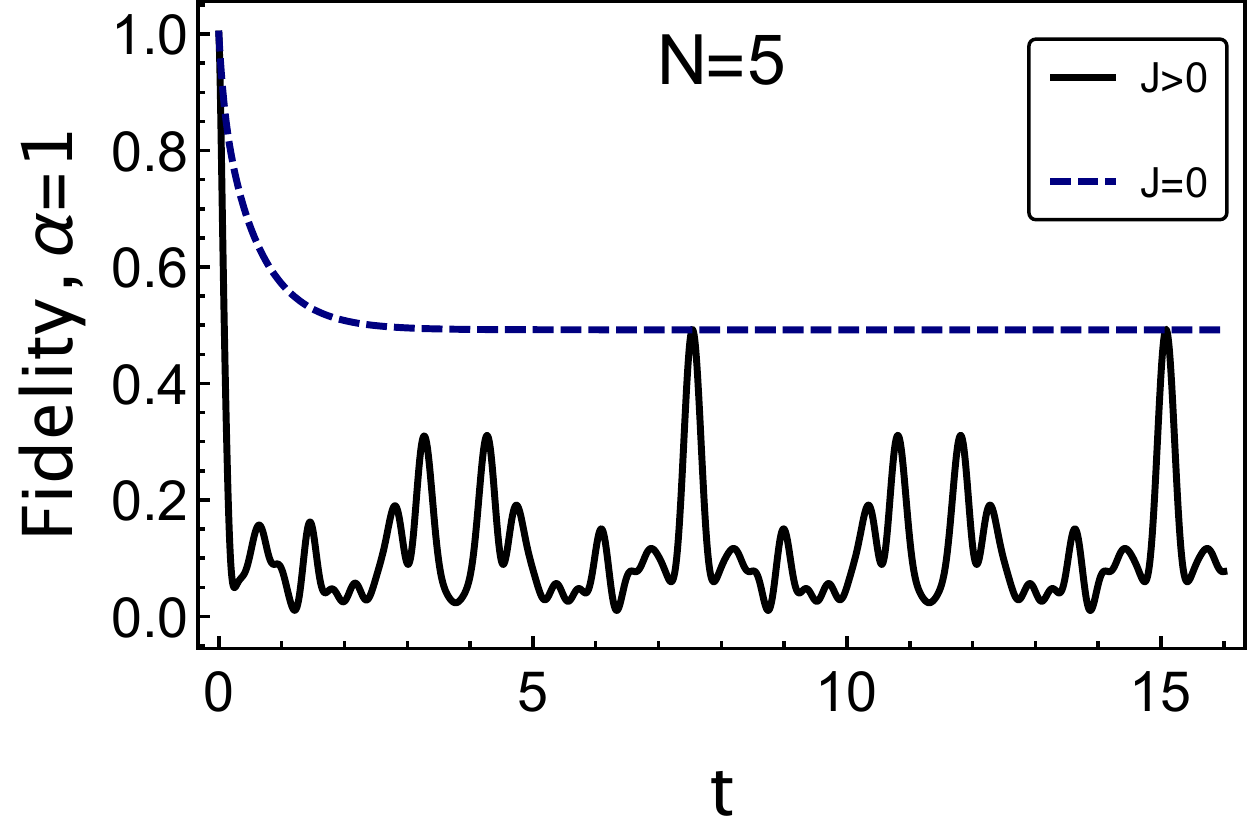}
\includegraphics[width=0.3\linewidth]{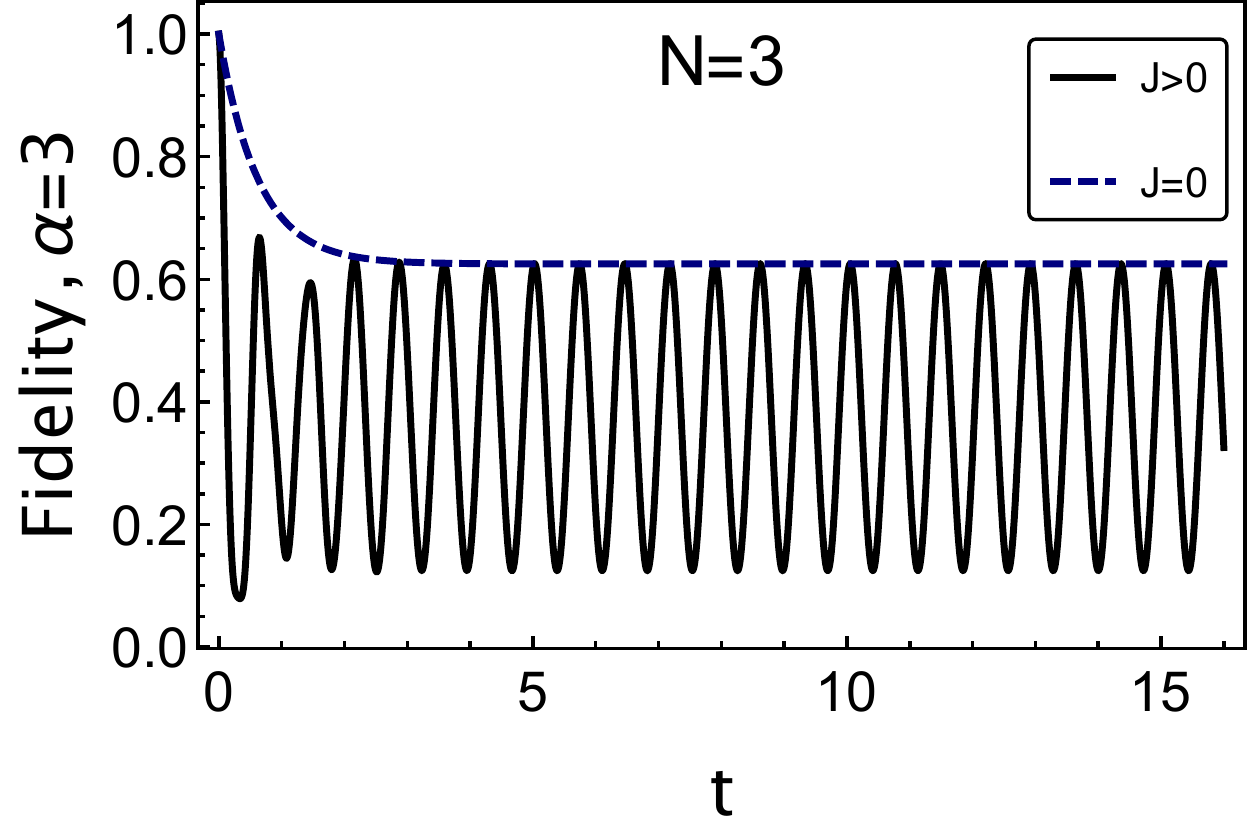}\ \ \ \includegraphics[width=0.3\linewidth]{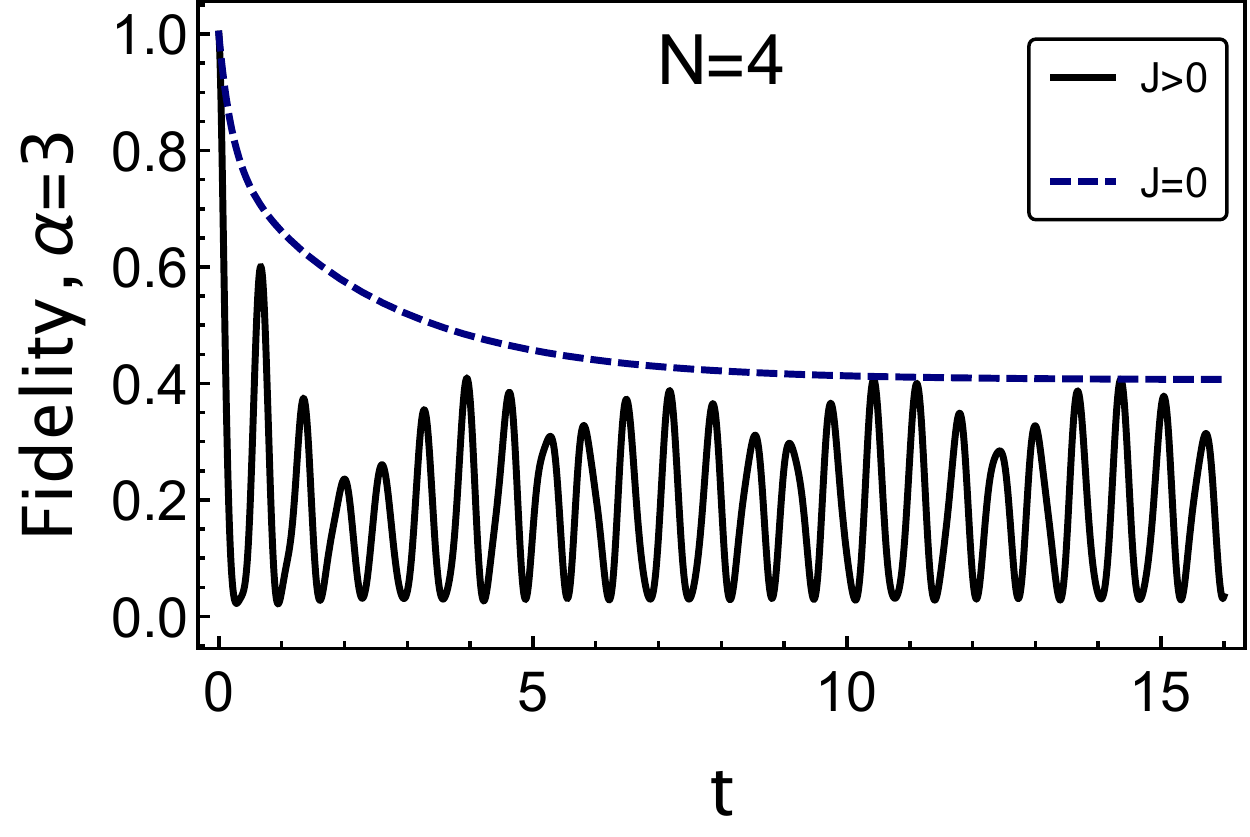}\ \ \ \includegraphics[width=0.3\linewidth]{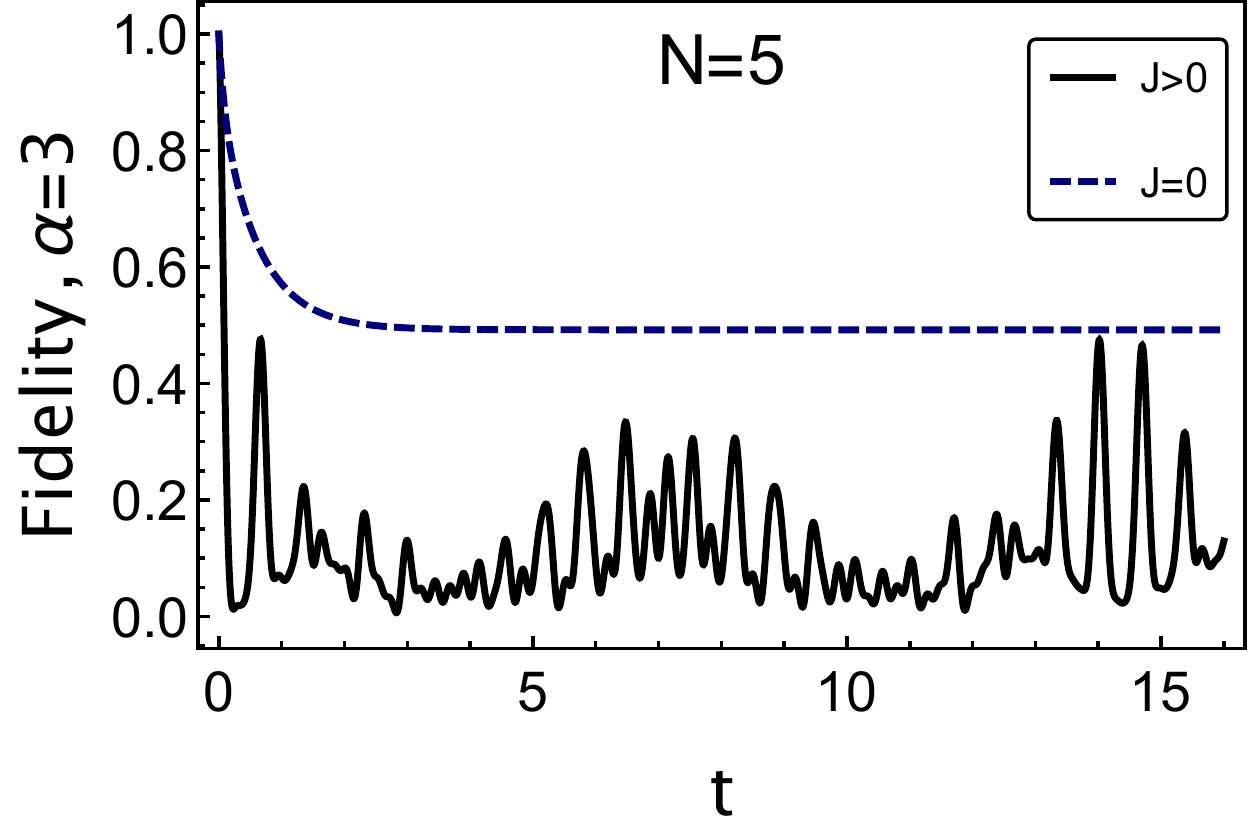}
\caption{\textbf{Fidelity of an initial product state for a dissipative long-range Ising model.} We show plots of the fidelity as a function of time for $N=3$ (on the left), $N=4$ (at the center), $N=5$ (on the right), with coupling constant $J_{ij}=J/|i-j|^\alpha$ for $\alpha=0$ (at the top), $\alpha=1$ (in the middle), $\alpha=3$ (at the bottom), with  $J=5$ (continuous curves) and $J=0$ (dashed curves). We fix the amplitude of the coupling constant with the environment $\gamma=0.2$.}
\label{SM:FigFidelity}
\end{center}
\end{figure}

\begin{figure}[t]
\begin{center}
\includegraphics[width=0.45\linewidth]{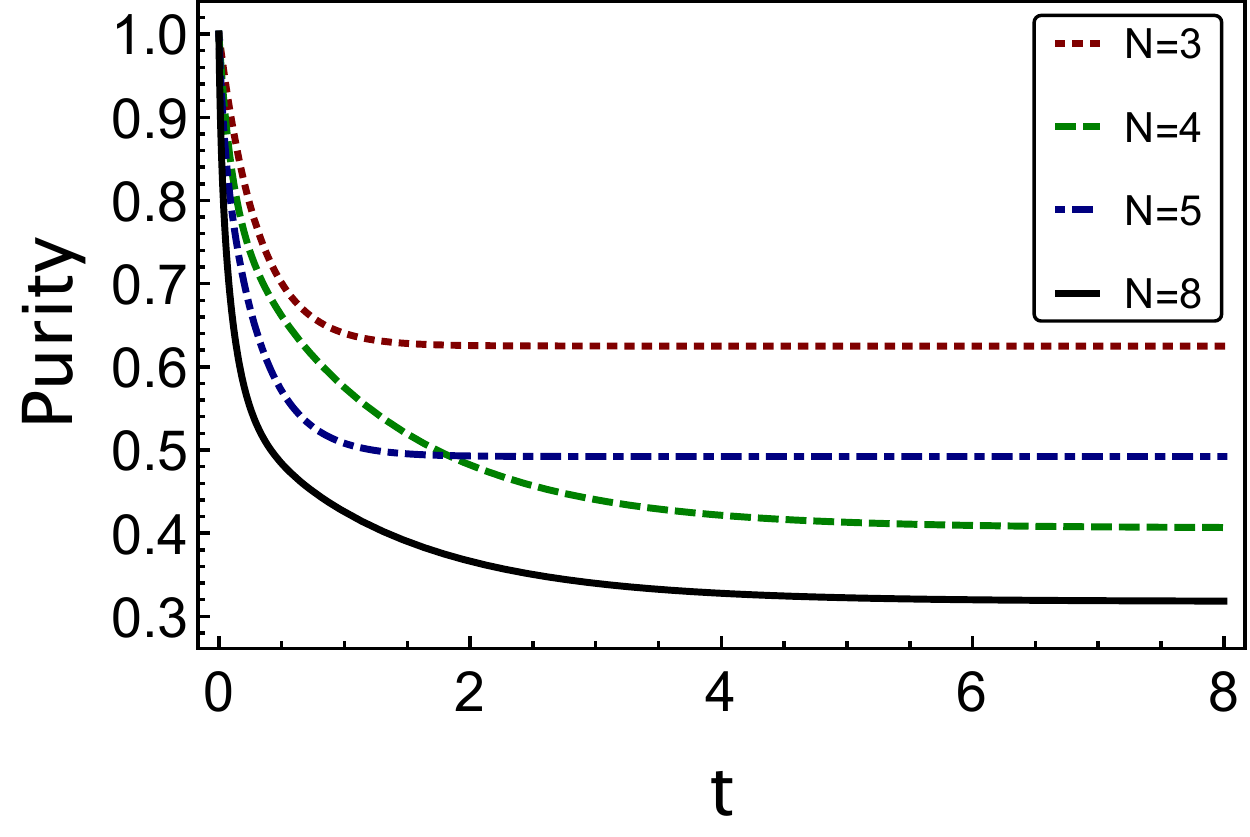}\ \ \includegraphics[width=0.45\linewidth]{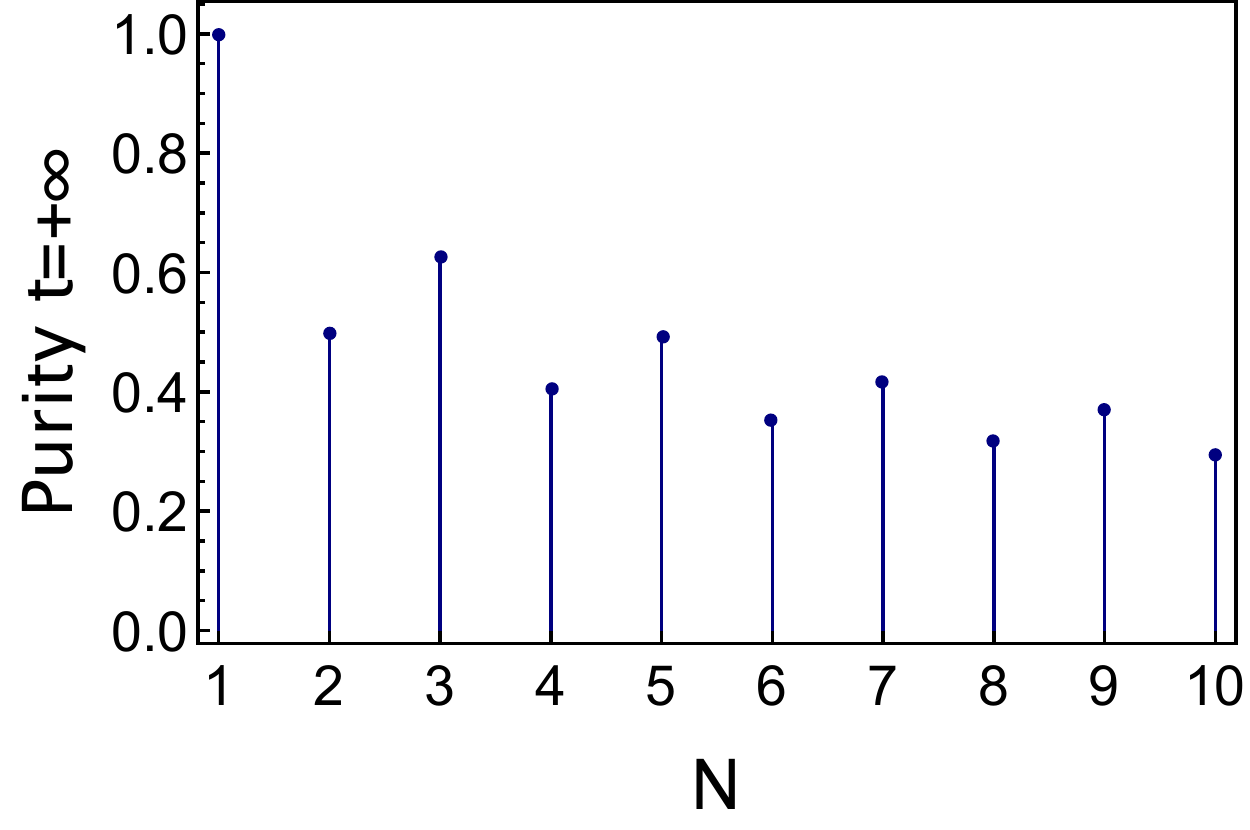}
\caption{\textbf{Purity of an initial product state for a dissipative long-range Ising model.} 
The purity as a function of time is plotted of the left side for $N=3,4,5,8$ with $\gamma=0.2$. At short time the purity decreases faster for larger number of particle. On the right side we show the asymptotic value of the purity obtained in the limit $t\rightarrow +\infty$.}
\label{SM:FigPurity}
\end{center}
\end{figure}

For instance, consider an initial product state $\rho(0)=\rho_0\equiv\bigotimes_{j=1}^{N}\rho_0^{(j)}$, where the $j$-spin initial density matrix is a pure state $\rho_0^{(j)}=|\phi\ra\la\phi|$ with $|\phi\ra = \frac{1}{\sqrt{2}}\left(|-1\ra+|1\ra\right)$. This state is pure and can also be written as $\rho_0 = |\Psi\ra\la\Psi|$ with $|\Psi\ra = \sum_{p=0}^{N}\sum_{m=1}^{{N\choose p}}|e_{p}^{(m)}\ra$. Then, the density matrix elements are given by $\rho_{IJ}(0)=2^{-N},\ \forall \{I,J\}$. Using equation \eqref{SM:fidelityInfty} and the properties of the spectrum of $L$ that we discussed above, we can show that the fixed point (defined only for pure dephasing model, i.e., for $\hat{H}_I=0$) is given by 
\begin{equation}\label{SM:fixstate}
\rho_{\infty} = \frac{1}{2^N}\sum_{p=0}^{N} {N \choose p} |D_p^N \ra\la D_{p}^N|\ , 
\end{equation}
where the Dicke state $|D_p^N \ra$ \cite{Dicke54} is the symmetrized version of the vector $|e_p^{(1)} \ra$, i.e., 
$$|D_p^N \ra= \frac{1}{\sqrt{{N \choose p}}}\sum_{m=1}^{{N \choose p}}|e_p^{(m)} \ra\ .$$ 
This state is also the fixed point of the Lindbladian operator with a mean-field-like interaction where $J_{ij}=J$ is constant for any couple of spins, as in this case the Hamiltonian is symmetric and proportional to the Lindblad operator $\hat{H}_I=J \hat{L}$ and so $[\hat{H}_I,\rho_{\infty}]=0$.

To illustrate our  results, Fig. \ref{SM:FigFidelity} shows the plot of the fidelity for $J_{ij}=J/|i-j|^\alpha$ with $\alpha=\{0,1,3\}$ and $N=\{3,4,5,8\}$. We also compare the time dependence of the purity  for different number of particles in Fig. \ref{SM:FigPurity}. Remarkably,  the purity at long time is surprisingly larger for $N=5$ than for $N=4$. This parity effect can be explained by the number of non-zero terms in the fixed point in equation \eqref{SM:fixstate}. For $N$ odd, one has $2\sum_{p=0}^{N}{N\choose p}^2=2{2N \choose N}$ terms equal to $1/2^N$ while we find $2\sum_{p=0}^{N}{N\choose p}^2-{N\choose N/2}^2=2{2N \choose N}-{N\choose N/2}^2$ terms for $N$ even. This leads to the asymptotic formula for the purity, see Eq.  \eqref{SM:purity}   
$$ p(\infty) = \frac{1}{2^{2N}}\left\{ \begin{array}{ll} 2{2N \choose N},\ \ \ \ \ \ \ \ \ \ \ \ \ N\ \text{odd}\ , \\ 2{2N \choose N}-{N\choose N/2}^2,\ N\ \text{even}\ ,\end{array} \right.$$
which decreases as a function of $2N$ and $2N+1$ and increases between two subsequent even and odd values.   

As for the decoherence time, the scaling is different for the initial product state $\rho_0$ than for an initial state $\rho_M$ defined in equation \eqref{SM:rhoM}. Indeed, by direct computation we find
\begin{align*}
\Delta_{\rho_0}L^2 &= \text{Tr}\left(L^2\rho_0\right) - \text{Tr}\left(L\rho_0\right)^2
=\sum_{i=1}^{2^N}l_i^2\rho_{ii}(0)
=\frac{1}{2^N}\sum_{p=0}^{N} {N\choose p}l_p^2
=\frac{N(N-1)}{2}\ ,
\end{align*}
where the matrix representation of the initial state is $\rho_{ij}(0)=2^{-N}$ leading to $\text{Tr}\left(L\rho_0\right)=0$ as $L=\sum_{i<j}\sigma_i\sigma_j$ is diagonal in the computational basis and as its trace vanishes. For large $N$, we find $\tau_D\sim 1/N^2$, which scales much larger than for $\rho_M$. 

\clearpage

\end{widetext}

\end{document}